\documentclass[twocolumn]{aastex631}

\usepackage{amsmath}
\usepackage{mathtools}
\usepackage{comment}

\begin{document}

\title{Prospects for gamma-ray emission from magnetar regions in CTAO observations }



\author[0000-0002-5438-3460]{M. F. Sousa}
\affiliation{Programa de Pós-Graduação em Física e Astronomia, Universidade Tecnológica Federal do Paraná (UTFPR), Av. Sete de Setembro, 3165, 80230-901, Curitiba, PR, Brazil}
\affiliation{Departamento de Engenharias e Exatas, Universidade Federal do Paraná (UFPR), Pioneiro, 2153, 85950-000 Palotina, PR, Brazil}

\author[0000-0002-2893-7589]{R. Jr. Costa}
\affiliation{Programa de Pós-Graduação em Física e Astronomia, Universidade Tecnológica Federal do Paraná (UTFPR), Av. Sete de Setembro, 3165, 80230-901, Curitiba, PR, Brazil}

\author[0000-0001-9386-1042]{Jaziel G. Coelho}
\affiliation{Programa de Pós-Graduação em Física e Astronomia, Universidade Tecnológica Federal do Paraná (UTFPR), Av. Sete de Setembro, 3165, 80230-901, Curitiba, PR, Brazil}
\affiliation{N\'ucleo de Astrof\'{\i}sica e Cosmologia (Cosmo-Ufes) \& Departamento de F\'isica, Universidade Federal do Esp\'irito Santo, 29075--910, Vit\'oria, ES, Brazil}
\affiliation{Divis\~ao de Astrof\'{\i}sica, Instituto Nacional de Pesquisas Espaciais, Avenida dos Astronautas 1758, 12227--010, S\~ao Jos\'e dos Campos, SP, Brazil}

\author[0000-0002-6463-2272]{R. C. Dos Anjos}
\affiliation{Programa de Pós-Graduação em Física e Astronomia, Universidade Tecnológica Federal do Paraná (UTFPR), Av. Sete de Setembro, 3165, 80230-901, Curitiba, PR, Brazil}
\affiliation{Departamento de Engenharias e Exatas, Universidade Federal do Paraná (UFPR), Pioneiro, 2153, 85950-000 Palotina, PR, Brazil}
\affiliation{N\'ucleo de Astrof\'{\i}sica e Cosmologia (Cosmo-Ufes) \& Departamento de F\'isica, Universidade Federal do Esp\'irito Santo, 29075--910, Vit\'oria, ES, Brazil}
\affiliation{Programa de pós-graduação em Física \& Departamento de Física, Universidade Estadual de Londrina (UEL), Rodovia Celso Garcia Cid Km 380, 86057-970 Londrina, PR, Brazil}
\affiliation{Programa de Pós-Graduação em Física Aplicada, Universidade Federal da Integração Latino-Americana, 85867-670, Foz do Igua\c{c}u, PR, Brazil}
\affiliation{Max-Planck-Institut für Kernphysik, Saupfercheckweg 1, D-69117 Heidelberg, Germany}

\begin{abstract}

Recent multi-wavelength observations have highlighted magnetars as significant sources of cosmic rays, particularly through their gamma-ray emissions. This study examines three magnetar regions—CXOU J171405.7-31031, Swift J1834-0846, and SGR 1806-20—known for emitting detectable electromagnetic signals. We assess the detectability of these regions using the upcoming Cherenkov Telescope Array Observatory (CTAO) by conducting an ON/OFF spectral analysis and compare the expected results with existing observations. Our findings indicate that CTAO will detect gamma-ray emissions from these three magnetar regions with significantly reduced emission flux errors compared to current instruments. In special, the study shows that the CXOUJ1714-3810 and SwiftJ1834-0846 magnetar regions can be observed by the full southern and northern CTAO arrays in just five hours of observation, with mean significances above $10 \,\sigma$ and $30 \,\sigma$, respectively. This paper discusses the regions analyzed, presents key results, and concludes with insights drawn from the study.
\end{abstract}



\section{Introduction} \label{sec:intro}

The search for galactic sources capable of accelerating particles up to the knee region and beyond has been a central focus of research and debate over the past few decades \citep{2023ApJS..264...50A, 2024JCAP...01..022A,2023EPJWC.28004001D,2023Galax..11...48C}. Potential accelerators include active galactic nuclei (AGNs), cosmological gamma-ray bursts (GRBs), supermassive black holes, tidal disruption events, millisecond pulsars/magnetars, and magnetar giant flares, both within and outside our Galaxy \citep{2013A&ARv..21...70B, 2017NCimC..40..126V,2023ARNPS..73..341C,2024PhRvL.132i1401P}. Supernova remnants (SNRs) located within massive stellar clusters and associated with molecular clouds are believed to be efficient accelerators, reaching energies up to $10^{15}$~eV \citep[][]{2019NatAs...3..561A, 2021JCAP...10..023D, Coelho_2022, 2023MNRAS.519..136V}. Recent discoveries by LHAASO\footnote{Large High Altitude Air Shower Observatory} and other gamma-ray studies have also pointed to the existence of PeVatrons within star-forming regions  \citep{2024ApJS..271...25C, 2019NatAs...3..561A, costa2024gamma}.

Gamma rays can shed light for studying several astrophysics processes and are a crucial instrument for analyzing the origin and acceleration of charged particles. These particles are mostly produced by accelerating and moving relativistic particles like protons or electrons. Through inverse Compton scattering, relativistic electrons can boost photons from the extragalactic background light (EBL) and cosmic microwave background (CMB) into the gamma-ray energy range. Additionally, gamma rays can result from hadronic cascades via pion decay \citep{2024ApJS..271...25C}. Multimessenger studies further enhance our understanding, with gamma rays providing an essential channel for probing the astrophysical origins of high-energy particles, as demonstrated in magnetar and other sources \citep{Sasse:2022pqn, Mocellin:2022occ, 2024MNRAS.531.3297K, 2010MNRAS.406L..25H, 2019FrASS...6...32H}.

Magnetars are neutron stars with extremely strong magnetic fields of $B\sim10^{14} - 10^{16}$ G, powered by the decay of the magnetic field \citep[][]{1992ApJ...392L...9D,1995MNRAS.275..255T}. While sources like AGNs, GRBs, and SNRs are often discussed, magnetars have emerged as significant contributors to cosmic ray acceleration, particularly through their gamma-ray emissions. Recent multimessenger observations have provided compelling evidence linking magnetars to these high-energy processes, making them key targets for further investigation \citep[see e.g.,][]{2010ApJ...725.1384H, 2011AdSpR..47.1317M, 2012ApJ...748...26K, 2018A&A...612A...1H}. These emissions have the potential to influence the observed levels of electrons, positrons, and cosmic rays detected on Earth \citep{2010MNRAS.406L..25H}.

In this research, we analyze three magnetar regions: CXOU~J171405.7–381031 \citep{2017ApJS..232...18A, 2023arXiv230712546B, 2022ApJS..260...53A, 2008A&A...486..829A}, Swift~J1834-0846 \citep{2015ApJS..218...23A, 2023arXiv230712546B,2022ApJS..260...53A,2006ApJ...636..777A}, and SGR~1806-20 \citep{2015ApJS..218...23A, 2023arXiv230712546B,2022ApJS..260...53A,2018A&A...612A..11H}, which emit electromagnetic counterparts detected by the Fermi Large Area Telescope \citep[Fermi-LAT;][]{Atwood2009} and the High Energy Stereoscopic System \cite[H.E.S.S.;][]{2018A&A...612A...1H}. The identification of acceleration regions and the discrimination between hadronic and leptonic mechanisms are only possible through precise measurements and data analysis. In this work, we demonstrate how the Cherenkov Telescope Array Observatory (CTAO) will enhance the future analysis of these regions through its advanced sensitivity capabilities. We apply the formalism presented by \cite{costa2024gamma} to assess the capability of the CTAO to detect gamma-ray emissions coming from magnetars regions, employing an ON/OFF spectral analysis approach through Gammapy\footnote{\url{https://gammapy.org/}} software \citep{2023A&A...678A.157D}. Gammapy is an open-source Python package designed for gamma-ray analysis, built on the foundations of Numpy\footnote{\url{https://numpy.org/}}, Scipy\footnote{\url{https://scipy.org/}}, and Astropy\footnote{\url{https://www.astropy.org/}}. Serving as a key component of the CTAO Science Analysis Tool, Gammapy offers a suite of tools for simulating the gamma-ray sky across various telescopes, including CTAO, H.E.S.S., Fermi-LAT, Very Energetic Radiation Imaging Telescope Array System \citep[VERITAS;][]{WEEKES2002221}, Major Atmospheric Gamma Imaging Cherenkov  \citep[MAGIC;][]{FERENC2005274}, and HAWC \citep{2014AdSpR..53.1492W}.

This study expands our understanding of magnetars and introduces methods and tools that can improve the precision and effectiveness of future observations. This paper is organized as follows. Section \ref{sec:2} provides details on the specific sources that were analyzed. Section \ref{sec:3} presents the modeling of the regions along with the anticipated performance of CTAO. The Gammapy results for the CTA spectrum simulation are presented in Section \ref{sec:4}, followed by the CTAO sensitivity analysis in Section \ref{sec:5}. In conclusion, Section \ref{sec:6} provides a summary of our findings.


\section{Magnetars Regions} \label{sec:2}

\subsection{CXOU J171405.7–381031}

CXOU J171405.7-381031 is a rapidly spinning magnetar located within the faint SNR CTB 37B \citep[][]{2008A&A...486..829A}. It was discovered through XMM-Newton observations. The pulsar has a spin period of $3.8$ seconds, and its spin down rate is estimated to be $\dot{P}=(6.40\pm 0.05)\times10^{-11}$ s/s \citep[][]{2010PASJ...62L..33S}. The derived parameters include a characteristic age of approximately $950$ yr, a magnetic field strength of $5\times 10^{14}$~G, and a spin-down luminosity of $4.5 \times 10^{34}$~erg/s.

What makes CXOU J171405.7-381031 particularly interesting is its status as the youngest known anomalous X-ray pulsar. Its age estimation aligns with the associated SNR thermal X-rays~\citep[see][]{2010ApJ...710..941H}. In addition, the TeV gamma-ray source HESS J1713-381 coincides with the region containing CXOU J171405.7-381031 \citep[][]{2006ApJ...636..777A}. While the TeV emission has been attributed to the SNR shell, the young age and rapid spin-down of CXOU J171405.7-381031 raise the possibility that the pulsar might also contribute to the TeV emission through inverse Compton scattering by a relic pulsar wind nebula (PWN); however, no direct X-ray evidence has been found to support the presence of a PWN contributing to this TeV emission \citep[see][for details]{2010ApJ...725.1384H}. 

Given the absence of non-thermal X-ray emission coinciding with TeV gamma-ray emission, \cite{2008A&A...486..829A} have suggested that the TeV source HESS~J1713-381 in this region likely originated from the decay of neutral pions produced in proton-nuclear interactions. However, observations $Suzaku$ detected both diffuse thermal and non-thermal X-ray emissions, with the non-thermal component being located in the western part of CTB 37B's radio shell \citep{2009PASJ...61S.197N}. This finding points to a multi-zone lepton scenario as a possible explanation for the gamma-ray emission. Furthermore, GeV emission detected by Fermi in the direction of CTB 37B has a spectrum that aligns smoothly with that of HESS~J1713-381, suggesting that the gamma-ray emission likely originates from a single source in this area \citep[see, e.g.,][]{2016ApJ...817...64X}.

\subsection{Swift J1834-0846}

Swift~J1834.9-0846 was discovered in 2011 during an outburst detected by the Swift Burst Alert Telescope (BAT) \citep[][]{2011GCN.12253....1D}. X-ray observations were carried out using various telescopes, including Swift, RXTE, Chandra (CXO), and XMM-Newton. RXTE and Chandra confirmed the magnetar nature of this object by measuring a spin period of $P = 2.48$ seconds and a period derivative of $\dot{P}= 7.96 \times10^{-12}$~s/s, indicating a surface magnetic field strength of about $10^{14}$~G \citep[see][]{2012ApJ...748...26K}. Located at a distance of approximately $4$~kpc, Swift~J1834.9-0846 resides at the center of the SNR W41, which has a radius of about $19$~pc \citep[see][and references therein]{2012ApJ...748...26K}. Notably, the SNR W41 is an order of magnitude larger than the X-ray nebula associated with Swift J1834.9-0846, which has a radius of approximately $2$~pc. The association with SNR W41 suggests a common age of around $5-100$ kyr, while the spin-down age of Swift J1834.9-0846 is estimated to be $4.9$ kyr \citep[see][]{2008AJ....135..167L}.

The region around Swift~J1834.9-0846 is a field rich in high-energy sources. This magnetar is located near the center of the extended TeV source HESS~J1834-087, which lies within the boundaries of the SNR~W41 \citep{2005Sci...307.1938A}. Additionally, Fermi-LAT detected an extended high-energy source, similar in size to the TeV source, coinciding with the same SNR \citep[see e.g.,][]{2023arXiv230712546B,2022ApJS..260...53A}. \cite{2012ApJ...748...26K} suggests that since the TeV source HESS~J1834-087 is significantly smaller than the SNR~W41, the TeV emission is unlikely to originate from the SNR shell. They also propose that the TeV emission could be powered by relativistic electrons injected by the magnetar formed after the SNR explosion. In contrast, \cite{2017MNRAS.464.4895G} argues that GeV/TeV emission is more likely to be of hadronic origin, resulting from the decay of neutral pions produced by the interaction of accelerated cosmic rays in the SNR shock with nuclei in a nearby giant molecular cloud.

\subsection{SGR 1806-20}

SGR 1806-20 is a magnetar, specifically classified as a soft gamma-ray repeater (SGR) in the constellation Sagittarius \citep[][]{2008MNRAS.386L..23B}. It has a rotation period of $P = 7.55$ seconds and a spin-down rate of $\dot{P}= 7.5 \times10^{-10}$~s/s \citep{2009PASJ...61S.387N}, indicating a spin-down power of approximately $10^{34}-10^{35}$~erg/s. This energy output appears inadequate to account for the quiescent X-ray emission with a luminosity of $10^{35}$~erg/s \citep[see][]{2014ApJS..212....6O}, which is more likely associated with the decay of its extremely strong magnetic field, theoretically estimated to be around $10^{15}$~G.

The soft X-ray emission ($E < 10$~keV) detected in SGR~1806-20 is typically explained by the presence of hot thermal gas, along with a non-thermal component resulting from inverse Compton scattering of thermal photons from the neutron star by electron/positron pairs in the neutron star's wind. For the hard X-ray emission ($E > 10$~keV), the origins are still debated, with possibilities including thermal bremsstrahlung, synchrotron radiation, and inverse Compton scattering \citep[see reviews by][]{2006RPPh...69.2631H,2011AdSpR..47.1317M}.

The emission mechanism of this pulsar also involves intense bursts of gamma rays resulting from magnetic field rearrangements within its crust. In particular, in 2004, SGR 1806-20 produced one of the most energetic gamma-ray bursts ever observed, briefly outshining the entire Milky Way \citep{2005Natur.434.1098H,2007ApJ...654..470W}. 

An extended very high-energy (VHE) gamma-ray source towards the magnetar SGR~1806-20 has been discovered by the H.E.S.S. telescopes \citep{2018A&A...612A..11H}. Named HESS~J1808-204, its emission characteristics include a VHE flux at $1$~TeV of a few $10^{-13}$ photons $\rm cm^{-2} s^{-1} TeV^{-1}$, with a photon index following a power-law spectrum. The luminosity, scaled to a distance of $8.7$~kpc, is approximately $1.6 \times 10^{34}$ $\rm erg s^{-1}$ in the energy range of $0.2$ to $10$~TeV . Interestingly, the VHE emission from HESS J1808-204 is extended, similar to the synchrotron radio nebula G10.0-0.3 associated with LBV~1806-20. The origin of the VHE emission remains unclear, but it could be related to the stellar wind processes of LBV~1806-20 or magnetic dissipation associated with SGR~1806-20. If HESS~J1808-204 is indeed associated with SGR 1806-20, the estimated young age of this magnetar \citep[ $\sim 650$~yrs;][]{2012ApJ...761...76T} suggests a limited diffusive transport distance of less than $30$~pc, which aligns with the observed size of the VHE emission region \citep[$15$~pc for a distance of $8.7$~kpc; see][]{2018A&A...612A..11H}.

\section{CTAO detectability} \label{sec:3}

In this section we investigate the ability of CTAO to observe VHE gamma-ray events in the vicinity of the magnetars CXOU J1714-3810, Swift J1834-0846, and SGR 1806-20. This is accomplished by employing two approximations from the study conducted by \cite{costa2024gamma}. First, a joint likelihood fit with data from Fermi-LAT
and H.E.S.S. instruments was used to determine the source's model in the multi-GeV to multi-TeV range (see Sec.~\ref{sec:CTA.1}). Furthermore, the 1D ON/OFF observation approach \citep{Piano2021} was employed to assess the anticipated performance of the CTAO by viewing the simulated source and determining the flux from the simulated observation (see Sec.~\ref{sec:CTA.2}).  

\subsection{Source modeling}\label{sec:CTA.1}
In order to determine the model of the source, we use data obtained from prior observations of electromagnetic counterparts located at a radius of less than $< 0.2^{\circ}$
from the position of the magnetar. The Table~\ref{tab:counterparts} shows the angular separation and celestial coordinates of each counterpart utilized in the joint likelihood fit. Two spectral models were evaluated to determine the shape of the source's spectrum. The exponential cutoff power law spectral model, described by

\begin{table*}
\centering
\caption{Celestial coordinates (RA, DEC.) and the angular separation (Sep.) of the counterparts present in the region of the three sources: CXOU J171405-381031 ($258.52 \deg$, $-38.17 \deg$), Swift J1834-0846 ($278.71 \deg$, $-8.76 \deg$) and SGR 1806-20 ($272.16 \deg$, $-20.41 \deg$).}

\begin{tabular}{lccc}
\hline
Counterpart & \multicolumn{1}{l}{\begin{tabular}[c]{@{}l@{}}RA \\ (deg)\end{tabular}} & \multicolumn{1}{l}{\begin{tabular}[c]{@{}l@{}}DEC\\ (deg)\end{tabular}} & \multicolumn{1}{l}{\begin{tabular}[c]{@{}l@{}}Sep.\\ (deg)\end{tabular}} \\ \hline
\multicolumn{4}{c}{CXOU J171405-381031} \\ \hline
3FHL J1714.0-381$^{a}$ & 258.50 & -38.20 & 0.03 \\
4FGL J1714.1-3811$^{b}$ & 258.53 & -38.19 & 0.01 \\
HESS J1713-381 (2008)$^{c}$ & 258.43 & -38.17 & 0.07 \\
HESS J1713-381 (2018)$^{d}$ & 258.46 & -38.22 & 0.06 \\ \hline
\multicolumn{4}{c}{Swift J1834-0846} \\ \hline
2FHL J1834.5-0846e$^{e}$ & 278.64 & -8.78 & 0.08 \\
3FHL J1834.5-0846e$^{a}$ & 278.64 & -8.78 & 0.08 \\
3FGL J1834.5-0841$^{f}$ & 278.63 & -8.68 & 0.12 \\
4FGL J1834.5-0846e$^{b}$ & 278.64 & -8.78 & 0.08 \\
HESS J1834-087 (2006)$^{g}$ & 278.71 & -8.74 & 0.02 \\
HESS J1834-087 (2018)$^{d}$ & 278.71 & -8.75 & 0.02 \\ \hline
\multicolumn{4}{c}{SGR 1806-20} \\ \hline
3FGL J1809.2-2016c$^{f}$ & 272.32 & -20.28 & 0.19 \\
4FGL J1808.2-2028e$^{b}$ & 272.05 & -20.48 & 0.13 \\
HESS J1808-204 (2018a)$^{d}$ & 272.17 & -20.40 & 0.01 \\
HESS J1808-204 (2018b)$^{h}$ & 272.00 & -20.40 & 0.15 \\ \hline
\end{tabular}%

\tablerefs{$^a$\citet{2017ApJS..232...18A}, $^b$\citet{2023arXiv230712546B,2022ApJS..260...53A}, $^c$\citet{2008A&A...486..829A}, $^d$\citet{2018A&A...612A...1H}, $^e$\citet{2016ApJS..222....5A}, $^f$\citet{2015ApJS..218...23A}, $^g$\citet{2006ApJ...636..777A}, $^h$\citet{2018A&A...612A..11H}.}

\label{tab:counterparts}
\vspace{0.4cm}
\end{table*}

\begin{equation}
\Phi(E) = \Phi_{0} \left ( \frac{E}{E_{0}} \right )^{-\Gamma } \mathrm{exp}\left ( -\lambda E \right ),
\label{eq:ecpl}
\end{equation}
where $\Phi$ is the flux as a function of energy $E$, $\Phi_{0}$ is the flux normalization at the reference energy $E_{0}$, $\Gamma$ is the spectral index, and $\lambda$ is the inverse of the $\gamma$-ray energy cutoff $E_{\mathrm{cut}}$. The second spectral model is the log parabola given by
\begin{equation}
\Phi(E) = \Phi_{0} \left ( \frac{E}{E_{0}} \right )^{-\Gamma -\beta \mathrm{log}\left ( \frac{E}{E_{0}} \right ) }, 
\label{eq:lp}
\end{equation}
where $\beta$ describes the spectral curvature.
Table \ref{tab:Spec_model} displays the best-fit parameters found for each magnetar region (see also Fig.~\ref{fig:p_flux}).
As a result, the observational data from the regions of CXOU J1714-3810 and Swift J1834-0846 were effectively characterized by the log parabola model and the exponential-cutoff power-law model, respectively. However, for the SGR 1806-20 magnetar region, no single model was sufficient to describe all the observations. Consequently, we separated the Fermi-LAT and H.E.S.S. observations and fitted each set individually using a log-parabola model.

\begin{table*}
\caption{Spectral model parameters were optimized using a simultaneous likelihood fit for datasets located in the region of each magnetar. The data from the CXOU~J1714-3810 and SGR~1806-20 magnetar regions were well-represented by the log parabola model, while the Swift J1834-0846 region showed a better fit with the exponential cutoff power-law model.}
\begin{tabular}{lccccc}
\hline
SOURCE & $\Gamma$ & $\Phi_{0}$ (cm$^{-2}$ s$^{-1}$ TeV$^{-1}$) & $E_{0}$ (TeV) & $\lambda$ (TeV$^{-1}$) & $\beta$ \\ \hline \hline
CXOU J171405-381031 & $2.482 \pm 0.08$ & $(8.90 \pm 0.73) \times 10^{-13}$ & 1.00 & - & $0.071 \pm 0.02$ \\
Swift J1834-0846 & $2.098 \pm 0.02$ & $(3.51 \pm 0.46) \times 10^{-14}$ & 10.00 & $0.184 \pm 0.07$ & - \\
SGR 1806-20 (Fermi-LAT) & $3.465 \pm 0.23$ & $(4.54 \pm 3.87) \times 10^{-15}$ & 1.00 & - & $0.062 \pm 0.018$ \\
SGR 1806-20 (H.E.S.S) & $2.583 \pm 0.17$ & $(1.10 \pm 0.42) \times 10^{-15}$ & 10.00 & - & $0.090$ \\ \hline
\end{tabular}%
\label{tab:Spec_model}
\vspace{0.4cm}
\end{table*}

\subsection{Expected performance of the CTAO}\label{sec:CTA.2}

Once we identified the most appropriate models for the datasets, we used the spectral parameters of these models to simulate CTAO observations using Gammapy \citep{2023A&A...678A.157D}. To achieve this, we employed the 1D ON/OFF observation technique for spectrum extraction \citep{Piano2021}. This commonly used technique in Cherenkov astronomy employs aperture photometry to measure emission from a source, such as counting the number of photons from a designated area. In this method, the ON region, which is centered on the source itself, is typically used to count the on-source photons (N$_{\rm{ON}}$). Additionally, one or more OFF regions are defined which possess the same characteristics as the ON region but lack significant $\gamma$-ray emission. These OFF regions are then used to count the off-source photons (N$_{\rm{OFF}}$), allowing for background estimation through the reflection method \citep[see e.g.,][]{2007A&A...466.1219B}. As a result, the estimated signal counts or photon excess can be calculated as $N_{S} = N_{\rm{ON}} - \alpha N_{\rm{OFF}}$, where $\alpha$ is the background efficiency ratio defined by \citep{Piano2021}

\begin{equation}
\alpha = \frac{A_{\rm{ON}} \cdot  t_{\rm{ON}}  \cdot  k_{\rm{ON}}}{ A_{\rm{OFF}} \cdot  t_{\rm{OFF}}  \cdot  k_{\rm{OFF}} }.
\label{eq:alpha}
\end{equation}
Here, $A$ represents the effective area, $t$ the exposure time, and $k$ the size of the region. Given that the ON region shares characteristics with the OFF regions, such as radius and offset from the center of the field of view, and by employing the reflected background method, Eq. (\ref{eq:alpha}) simplifies to $\alpha = 1/N$, where $N$ is the number of OFF regions. In our simulation, we used an alpha value of $0.1$, and the detection significance is calculated as \cite{1983ApJ...272..317L}:

\begin{equation}
\begin{split}
S &=  \sqrt{2} \Bigg\{ N_{\rm{ON}} \, \rm{ln}\left [ \frac{1+\alpha }{\alpha} \left ( \frac{N_{\rm{ON}}}{N_{\rm{ON}} + N_{\rm{OFF}} } \right ) \right ] \\
&\quad+ N_{\rm{OFF}} \, \rm{ln}\left [ (1+\alpha) \left ( \frac{N_{\rm{OFF}}}{N_{\rm{ON}} + N_{\rm{OFF}} } \right ) \right ]  \Bigg\}^{1/2}.
\end{split}
\label{eq:LIandMA}
\end{equation}

The value of $N_{\rm{OFF}}$ is estimated through the background template stored in the instrument response functions (IRFs - version prod5 v0.1) from \cite{CTA_observatory_2021_5499840}. These IRFs include all of the dependencies of performance parameters and outline the anticipated capabilities of the instrument. They have been calculated for both the proposed southern and northern arrays, as well as for various telescope sub-arrays observing an object at three distinct zenith angles ($z$): $20^{\circ}$, $40^{\circ}$, and $60^{\circ}$. Each IRF was analyzed using different criteria, considering observation times of $t_{\rm obs} = (0.5, , 5.0, , 50.0)$ hours. The prod5 v0.1 IRFs assume that the CTAO arrays are configured as Alpha, with 4 Large-Sized Telescopes (LSTs) and 9 Medium-Sized Telescopes (MSTs) in the northern array, and 14 MSTs and 37 Small-Sized Telescopes (SSTs) in the southern array. In our study, we utilized various IRFs corresponding to both the northern and southern arrays, along with their sub-arrays, across the three zenith angles ($20^{\circ}$, $40^{\circ}$, $60^{\circ}$). Additionally, we considered that all IRFs are azimuth averaged, as recommended in the CTAO IRF documentation for general studies \citep[see][]{CTA_observatory_2021_5499840}.

\vspace{1cm}
\section{ CTAO spectrum simulation} \label{sec:4}

We performed simulations of CTAO observations using various IRF configurations, covering $t_{\rm obs} =$ ($5.0$, $10.0$, $30.0$, $50.0$, $100.0$)~hours. The goal was to determine the optimal telescope sets and observational parameters for very high-energy sources located within magnetar regions.

To simulate a specific observation, the IRFs are reduced to an appropriate geometry and combined with a source model to forecast the expected photon count rates for each energy bin. These predicted counts are then used to generate a simulated dataset using a Poisson probability distribution. For each magnetar region, three thousand ON-OFF spectral observations were simulated, based on the spectral models detailed in Table~\ref{tab:Spec_model}. Specifically for the SGR~1806-20 source, the spectral model corresponding to the H.E.S.S. data was used, as it falls within the energy range intended for the CTAO. The ON region was defined as a circular area with a radius of $0.2^{\circ}$, centered on the magnetar's location, aligned with the counterpart analysis region. The telescope's pointing position was similarly adjusted to maintain a fixed azimuthal offset of $0.5^{\circ}$ from the source's position. To account for variations in point spread function (PSF) leakage, the integration radius was treated as energy-dependent, employing a constant flux enclosure proportion 68\% to adjust the background estimation for the size of the region varying with energy.

\begin{figure*}[htb]
\centering
\gridline{\fig{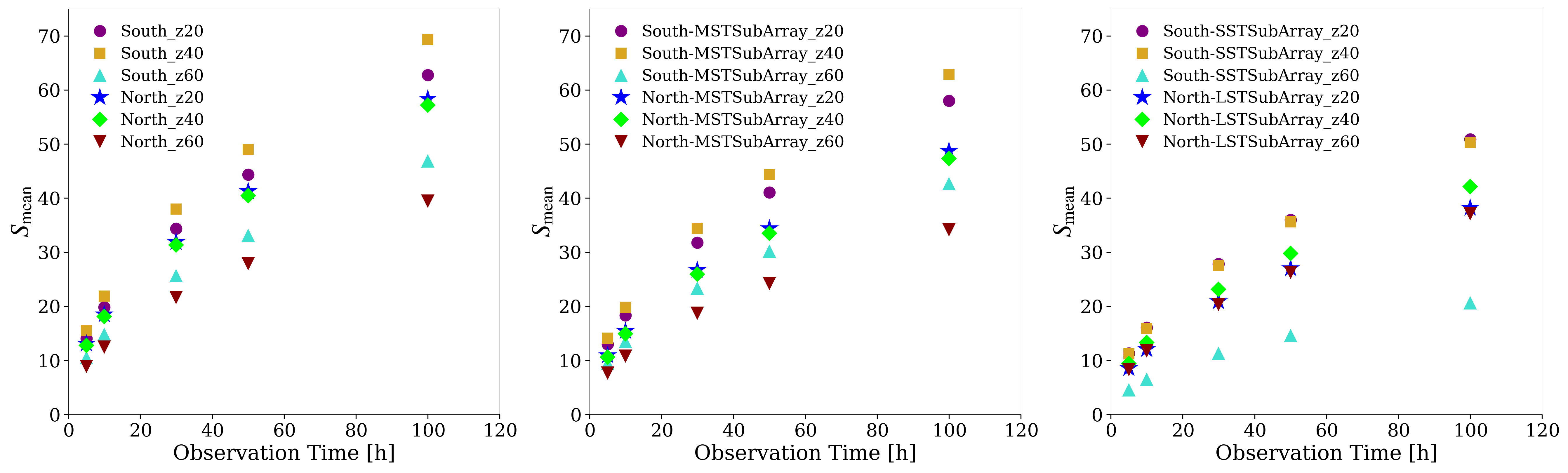}{\textwidth}{(a) CXOU~J1714-3810 region}}
\gridline{\fig{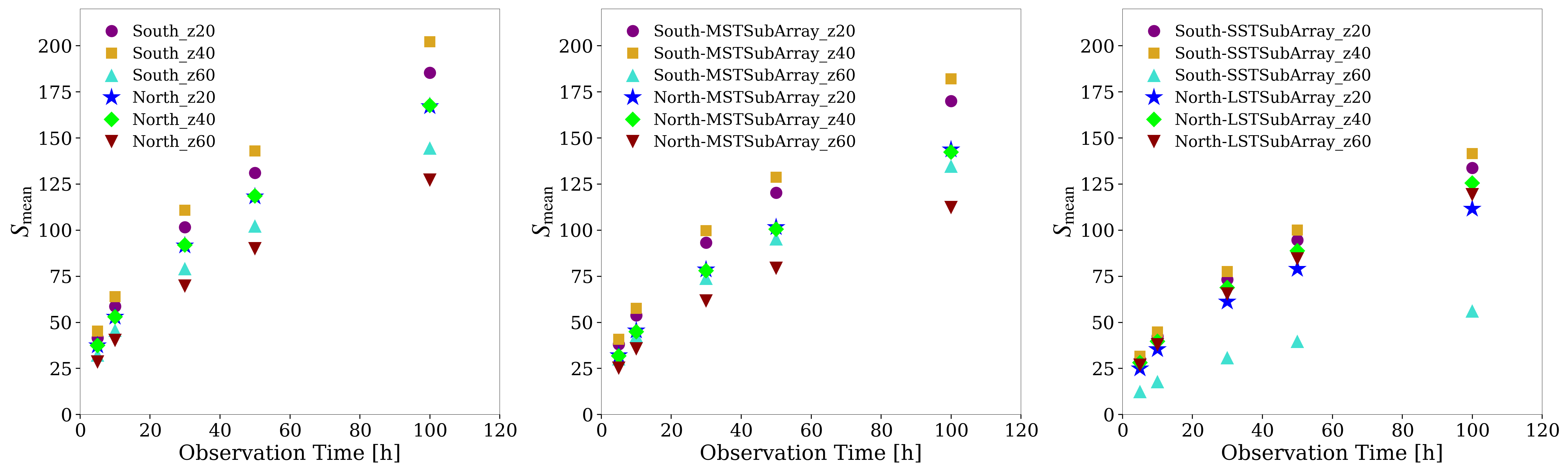}{\textwidth}{(b) Swift~J1834-0846 region}}
\gridline{\fig{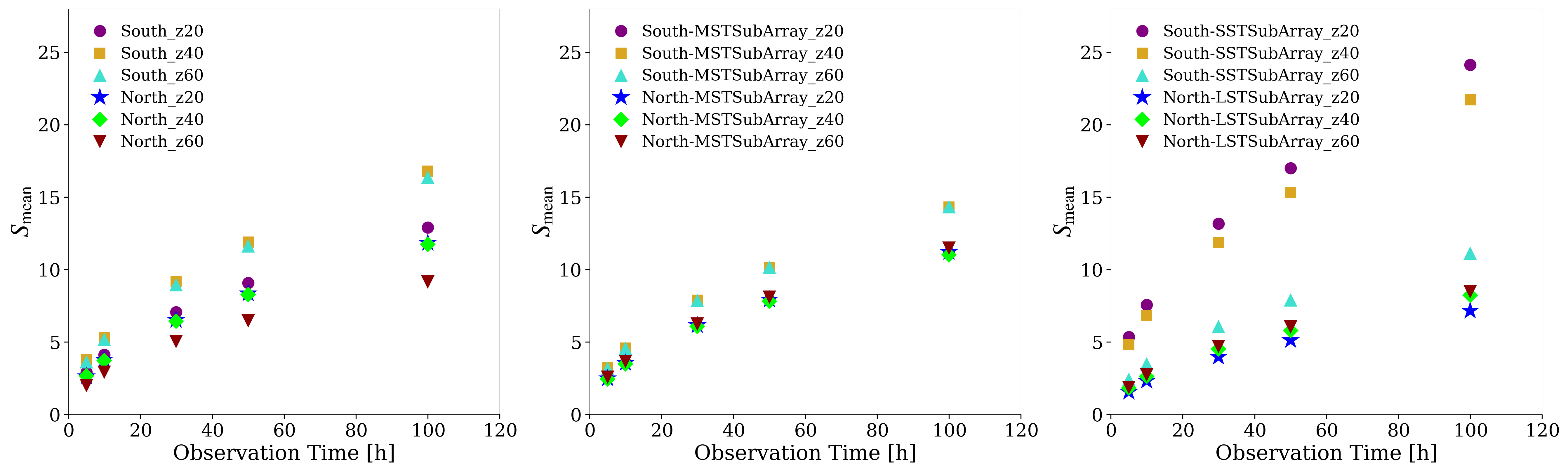}{\textwidth}{(c) SGR~1806-20 region}}
\caption{Mean statistical significance as a function of observation times (5h, 10h, 30h, 50h, and 100h) for magnetar regions. $S_{\rm mean}$ indicates the average value of $S$, calculated over 3000 simulated spectra, based on the spectral model detailed in Table~\ref{tab:Spec_model}. The North and South arrays refer to the use of all telescopes at their respective sites, while the sub-arrays pertain to telescopes of the same size within either the North or South location. Furthermore, "z20," "z40," and "z60" correspond to zenith angles of $20^{\circ}$, $40^{\circ}$, and $60^{\circ}$, respectively, as incorporated into the IRF performance estimates.}
\label{fig:gamma-models}
\end{figure*}

Figure \ref{fig:gamma-models} presents the average statistical significance values ($S_{\rm mean}$, see Eq. (\ref{eq:LIandMA})) of the simulated ON-OFF spectra, calculated across 3000 iterations and plotted against observation time for CXOU~J1714-3810, Swift~J1834-0846, and SGR~1806-20, respectively. These figures illustrate different configurations of the CTAO array based on the IRFs employed. The North and South arrays represent the use of all telescopes at their respective sites for observations, while the North and South sub-arrays are categorized by telescope size, with LST and MST sub-arrays in the North and MST and SST sub-arrays in the South. Additionally, the figures display results for three distinct zenith angles: $20^{\circ}$, $40^{\circ}$, and $60^{\circ}$.

In Fig.\ref{fig:gamma-models}-(a), it is evident that the CXOU~J1714-3810 region can be observed by both the CTAO's full southern and northern arrays, achieving a $S_{\rm mean} \gtrsim 10 \, \sigma$ for observation times exceeding 5 hours. Additionally, due to the source's position in the sky, the IRFs for the southern full array at $z = 40^{\circ}$ show superior performance in detecting the modeled source, with significance values reaching approximately $S_{\rm mean} \sim (22, \, 38,  \, 50,  \, 70) \sigma$ for observation times of 10, 30, 50, and 100 hours, respectively. When observations are conducted using only the sub-arrays, a decrease in significance is noted compared to full arrays at the same location and zenith angle. This decrease occurs because the total detected counts depend on the source model, the background counts, and the type of telescope used. Each type of telescope is optimized for different energy ranges: LSTs are sensitive to CTAO's lower energies ($\sim 20 - 150$~GeV), MSTs to medium energies ($\sim 150$~GeV$- 5$~TeV), and SSTs to higher energies ($\sim 5 - 300$~TeV) \citep[see][]{KSP}. Consequently, in this region, signal counts (excess) observed by the CTAO are higher in the mid-energy range compared to the high- and low-energy ranges.

The analysis for the Swift~J1834-0846 region, shown in Fig.\ref{fig:gamma-models}-(b), reveals trends similar to those observed in the CXOU~J1714-3810 region, but with significantly higher significance values. With observation times exceeding 5 hours, the CTAO's southern and northern arrays can detect the modeled source region with a significance of $\gtrsim 30 \, \sigma$. Specifically, the significance values are about $S_{\rm mean} \sim (65, \, 110, \, 145, \, 200) \sigma$ for observation times of 10, 30, 50, and 100 hours, respectively, when using the IRFs of the full South array configured at $z = 40^{\circ}$. These IRFs show the best performance among the CTAO configurations analyzed. Among the MSTs sub-arrays, as well as when comparing both SSTs and LSTs sub-arrays, the South location at a zenith angle of $z = 40^{\circ}$ also yields the highest significance. Figure \ref{fig:gamma-models}-(b) illustrates the best CTAO configuration for the Swift~J1834-0846 region, considering both the desired detection significance and the energy range of interest.

The statistical significance of CTAO observations in the SGR~1806-20 region (Fig.~\ref{fig:gamma-models}-(c)) indicates a lower photon flux from this source compared to the previous regions. Depending on the selected IRFs, significance values can vary between $S_{\rm mean}\sim 2 \, \sigma$ and $S_{\rm mean}\sim 24 \,\sigma$ for observation times ranging from 5 to 100 hours. Unlike the other magnetar regions, we found that for SGR~1806-20, a sub-array rather than a full array offers the best detection performance: specifically, the SSTs sub-array at $z = 20^{\circ}$. This is because the modeled source shows a higher excess count at the highest energies, where the SSTs sub-array's sensitivity is optimized. In the CTAO's low- and mid-energy ranges, most signal counts fall below background levels, which are particularly high at lower energies and decrease with higher energies. Excess counts begin to appear at approximately 300 GeV. Consequently, considering the detected N$_{\rm{ON}}$ and N$_{\rm{OFF}}$ for each set of IRFs, it is logical that the SSTs sub-array exhibits greater significance (see Eq. (\ref{eq:LIandMA})). However, it is worth noting that although this configuration provides the highest statistical significance for detecting the SGR 1806-20 region, its use limits the analysis of source information to the highest energy range of the CTAO.

\section{CTAO sensitivity} \label{sec:5}

We can further assess the detectability of the CTAO by examining the sensitivity curves configured for a fixed zenith angle and offset. In this approach, significance is determined through a 1D analysis (ON-OFF regions) and calculated using the Li$\&$Ma formula. Unlike simulations of CTAO observations, where detection significance is computed based on the total simulated counts across the entire energy range optimized for each IRF set, sensitivity curves are produced by setting a minimum significance value per bin, a minimum expected signal count per bin, and ensuring the excess exceeds a specific background threshold. This allows us to determine the minimum detectable flux value for each bin on the basis of one of these three criteria.

\begin{figure*}
    \centering
    \resizebox{5.9cm}{!}{\includegraphics{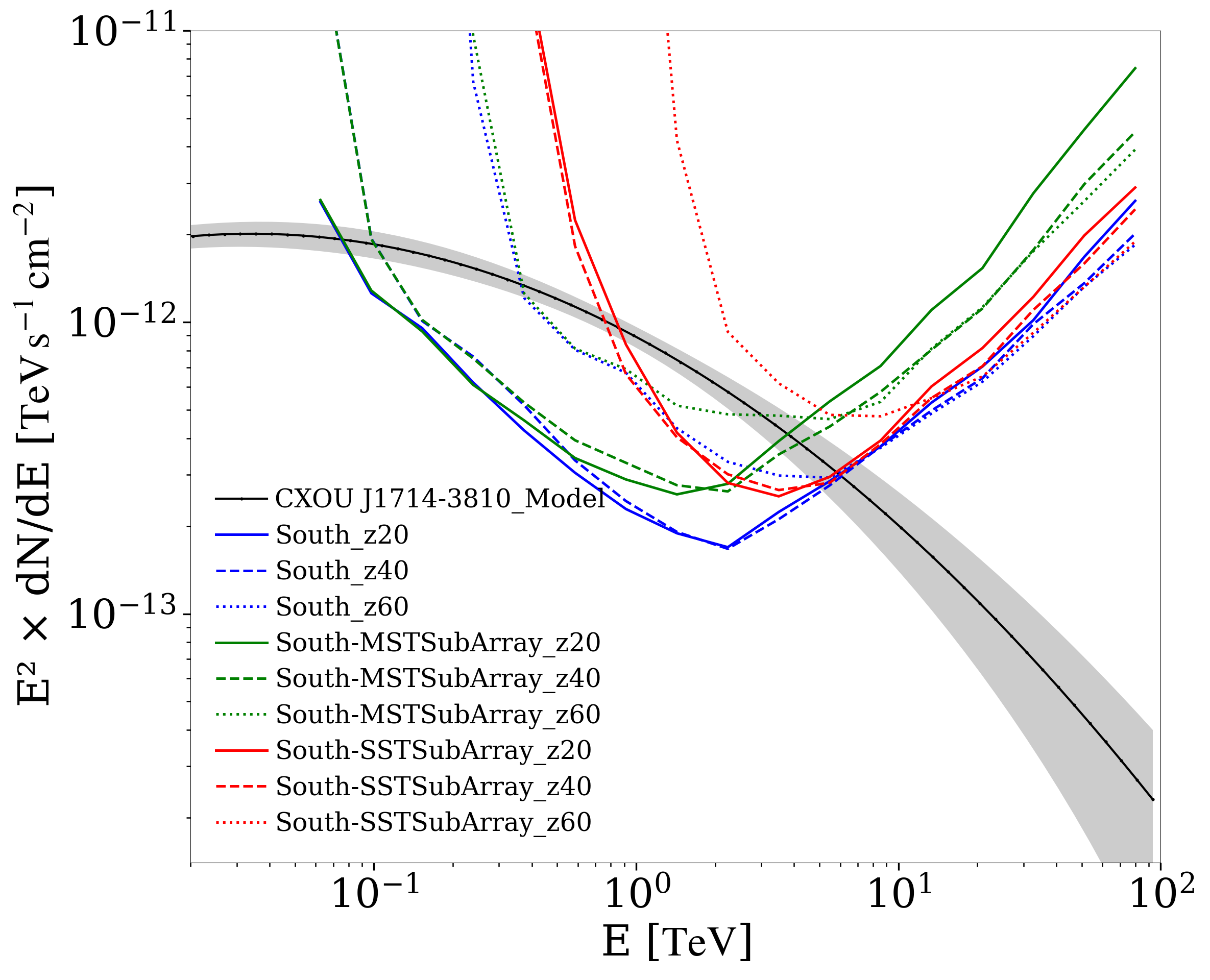}}
    \resizebox{5.9cm}{!}{\includegraphics{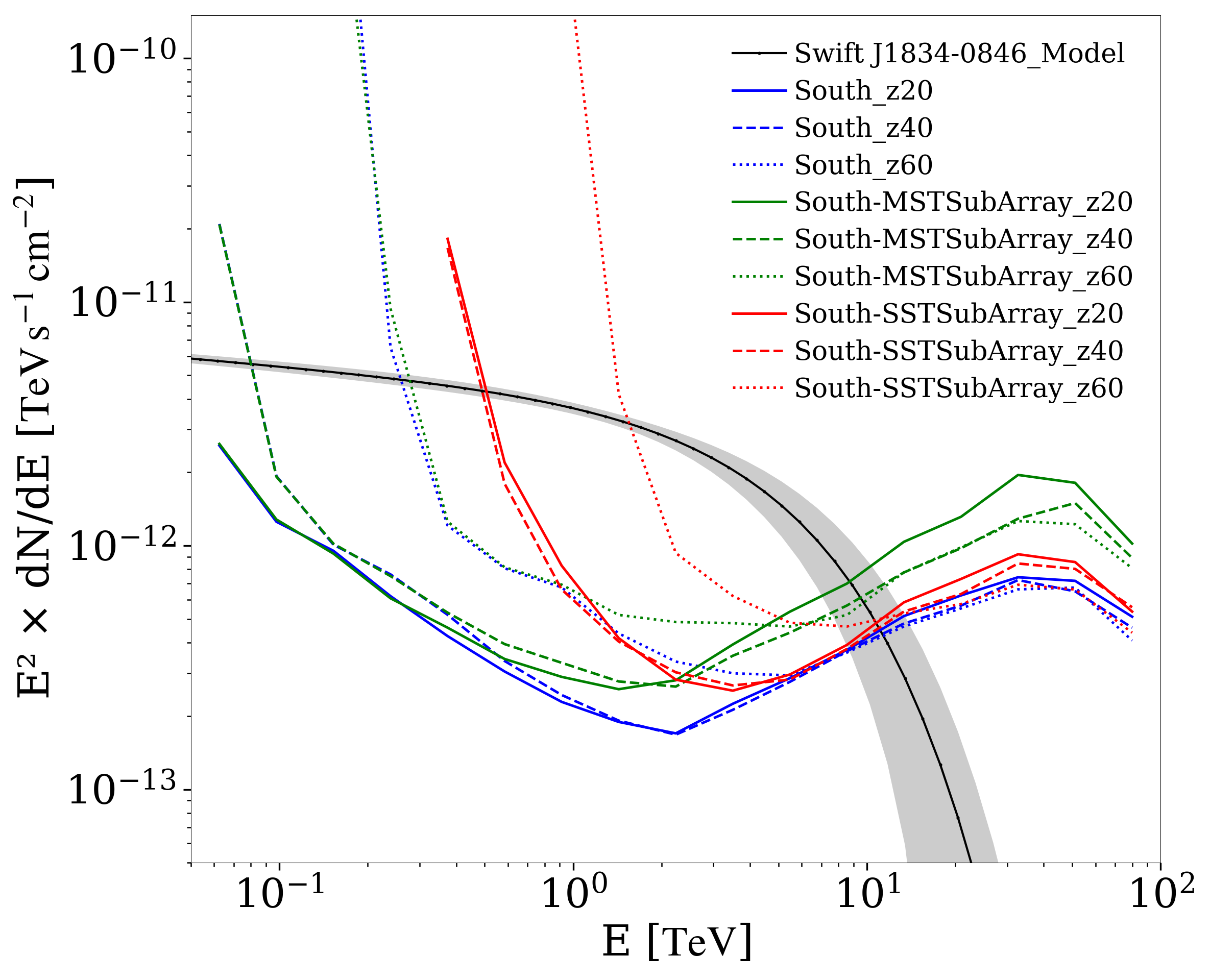}}
    \resizebox{5.9cm}{!}{\includegraphics{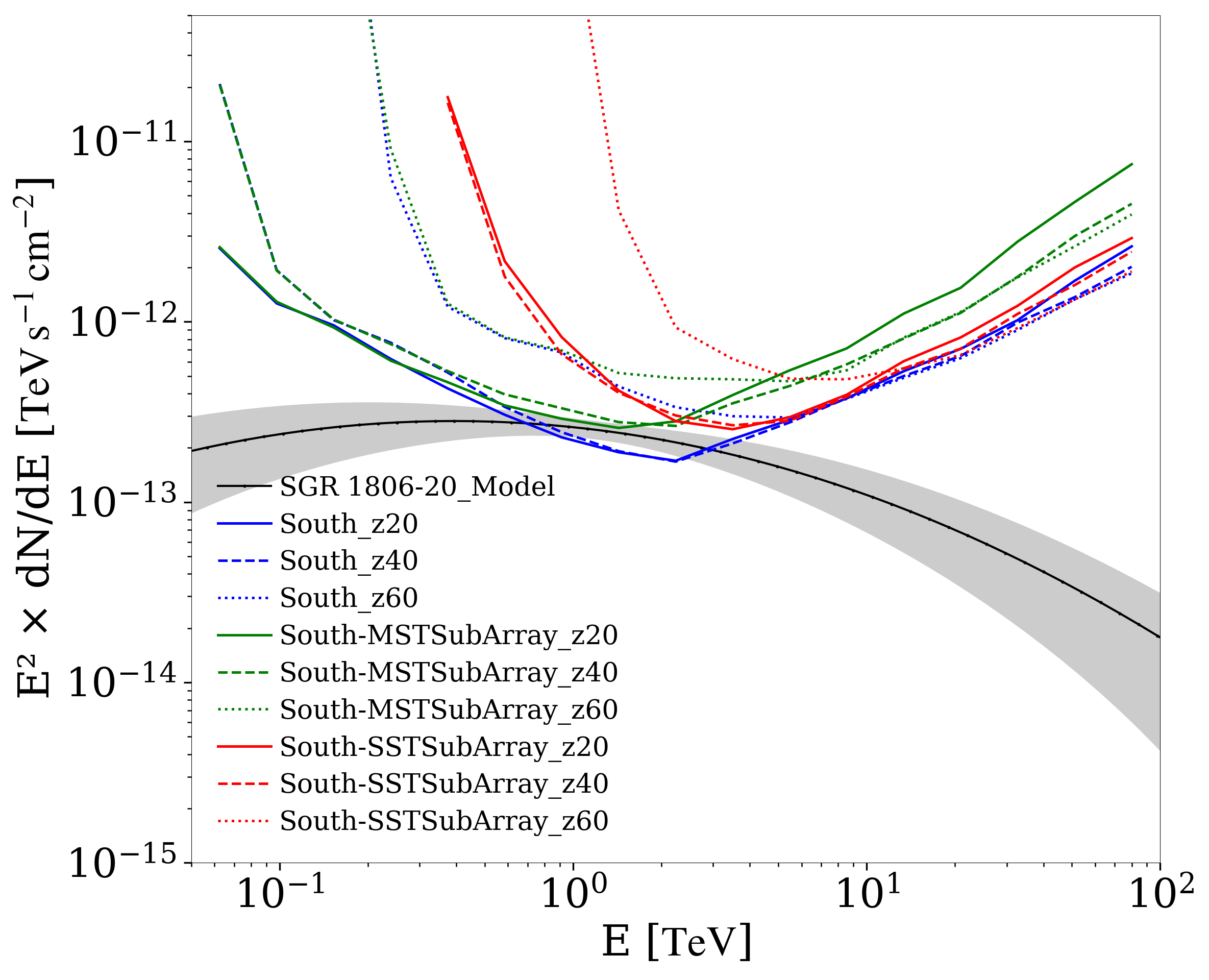}}
    \resizebox{5.9cm}{!}{\includegraphics{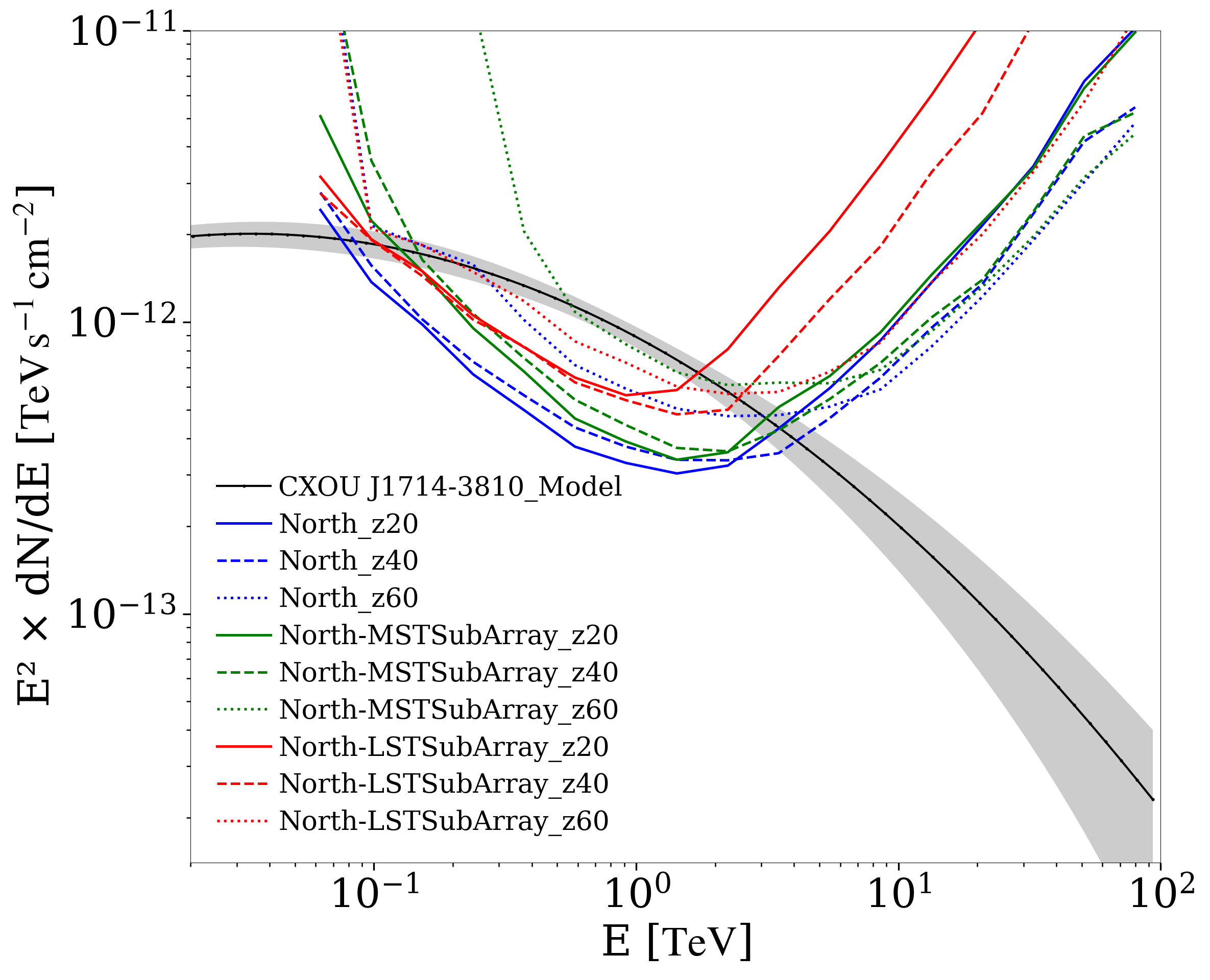}}
    \resizebox{5.9cm}{!}{\includegraphics{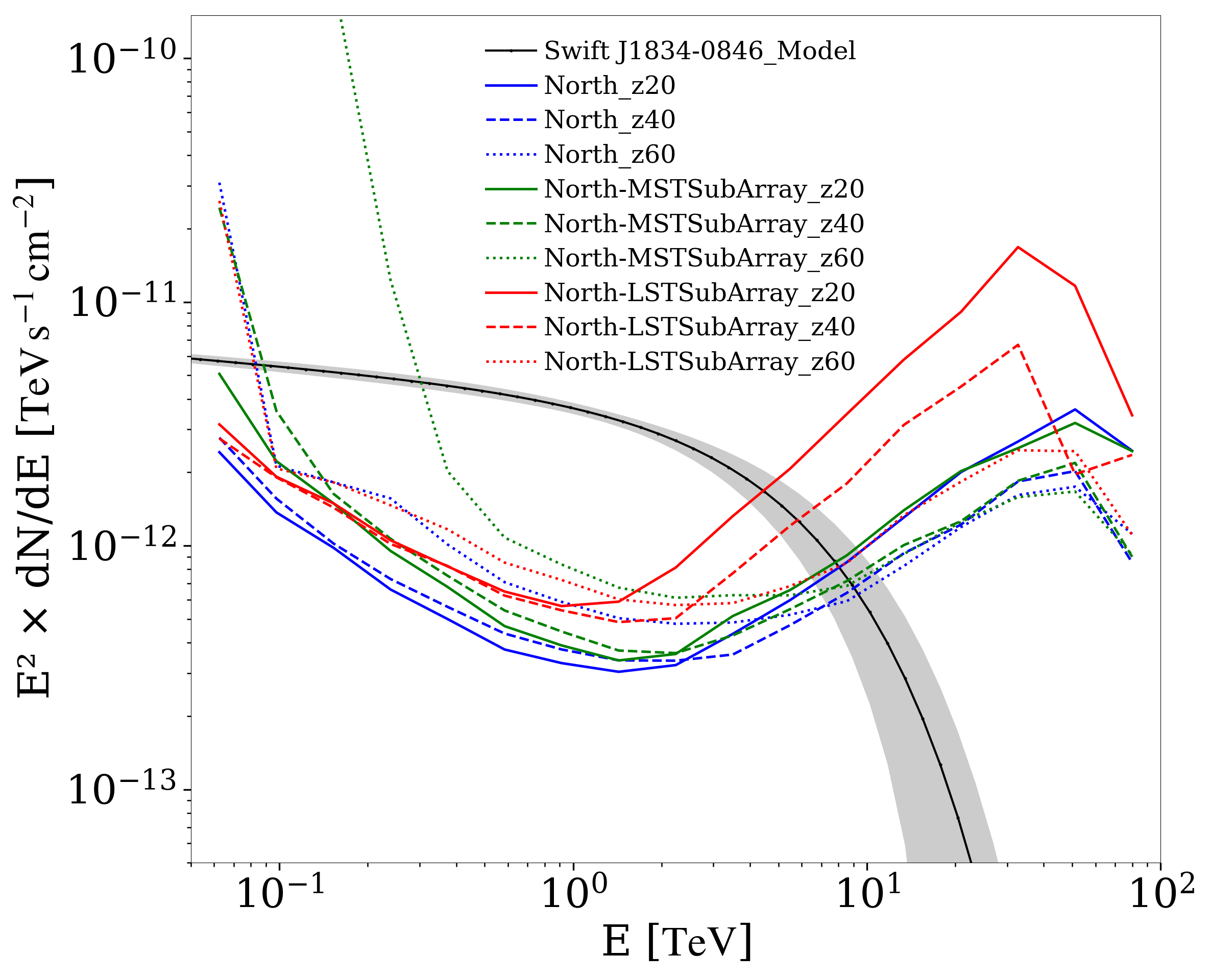}}
    \resizebox{5.9cm}{!}{\includegraphics{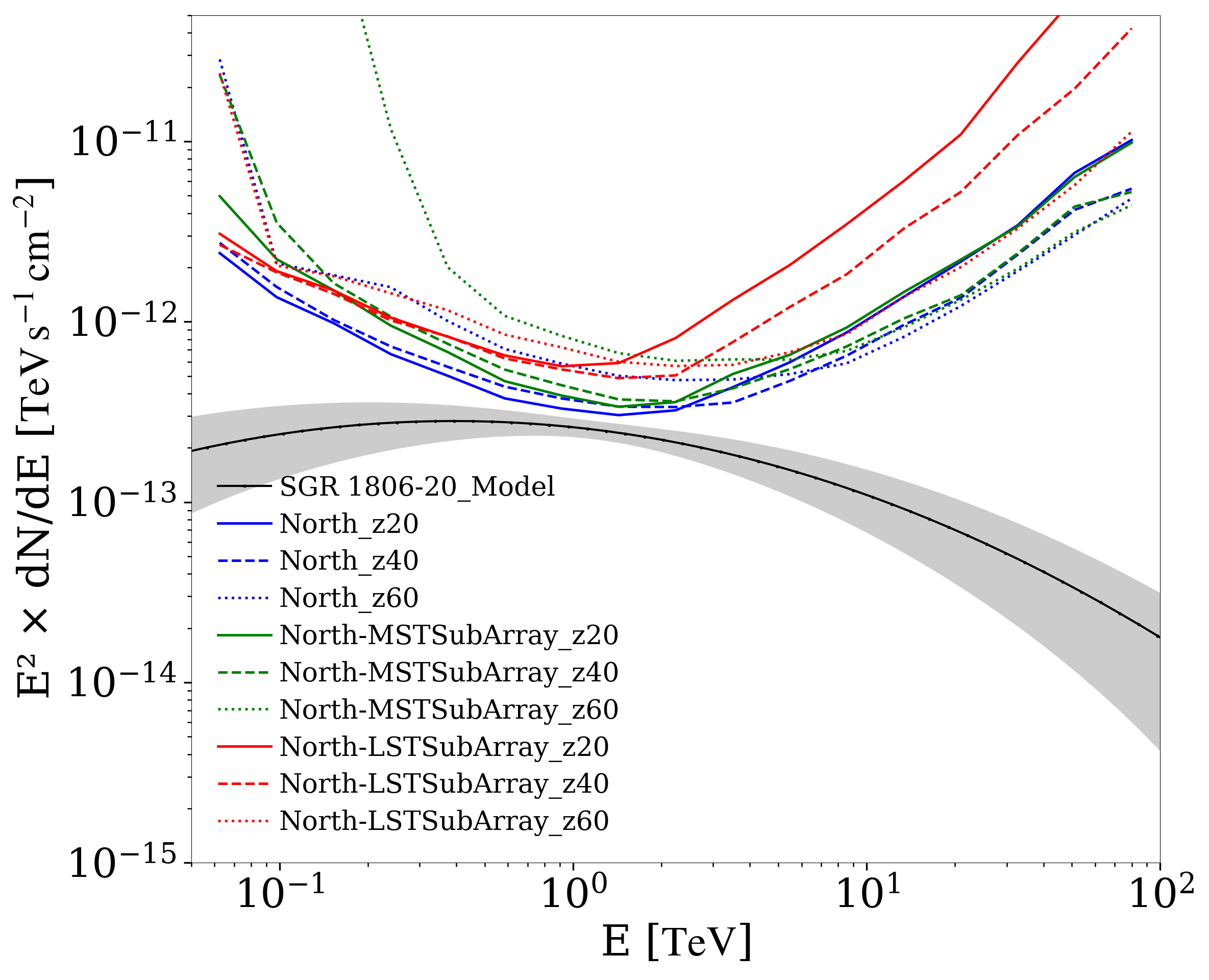}}
    \caption{The source spectral energy distribution and the CTAO's differential sensitivity curves are presented for the North and South arrays, sub-arrays with various zenith angles. The right, middle, and left panels illustrate the spectra of the CXOU~J1714-3810, Swift~J1834-0846, and SGR~1806-20 magnetar regions, respectively. The upper panels display the sensitivity curves for the Southern arrays and sub-arrays, while the lower panels show those for the Northern arrays and sub-arrays.}
    \label{fig:sens_10h}
    \vspace{0.3cm}
\end{figure*}

Using the same configuration previously defined for the ON and OFF regions and the pointing position offset, we generated sensitivity curves based on the same set of IRFs used in the CTAO observation simulations. For this, we established thresholds of a minimum statistical significance of 5$\sigma$, a minimum detected gamma-ray count of 10, and a signal-to-background ratio of $5\%$ in each energy bin, which are standard conditions for CTAO sensitivity calculations. Flux sensitivity was evaluated over an energy range from $500$~GeV to $100$~TeV, using five logarithmic bins per decade. Additionally, consistent with previous considerations, we implemented an energy-dependent integration radius with a $68\%$ containment correction to account for PSF variation. Figure \ref{fig:sens_10h} presents the CTAO sensitivity curves for a 10-hour observation time, along with the fitted spectral models (see Table \ref{tab:Spec_model}) for the three magnetar regions.

Based on established criteria for flux sensitivity, Figure~\ref{fig:sens_10h} shows that all CTAO configurations examined are capable of capturing at least part of the spectra for the CXOU~J1714-3810 and Swift~J1834-0846 magnetar regions. However, the IRFs of the South SST sub-array optimized for $z = 60^{\circ}$ lack sufficient sensitivity to detect the CXOU~J1714-3810 region. Concerning the SGR 1806-20 region, neither the north arrays and subarrays nor the south subarrays are able to detect the flux from this source. Furthermore, the full South arrays at $z = 20^{\circ}$ and $z = 40^{\circ}$ only capture a small portion of the spectrum, suggesting that longer observation times are necessary for detection.

\begin{figure}
    \includegraphics[width=\hsize,clip]{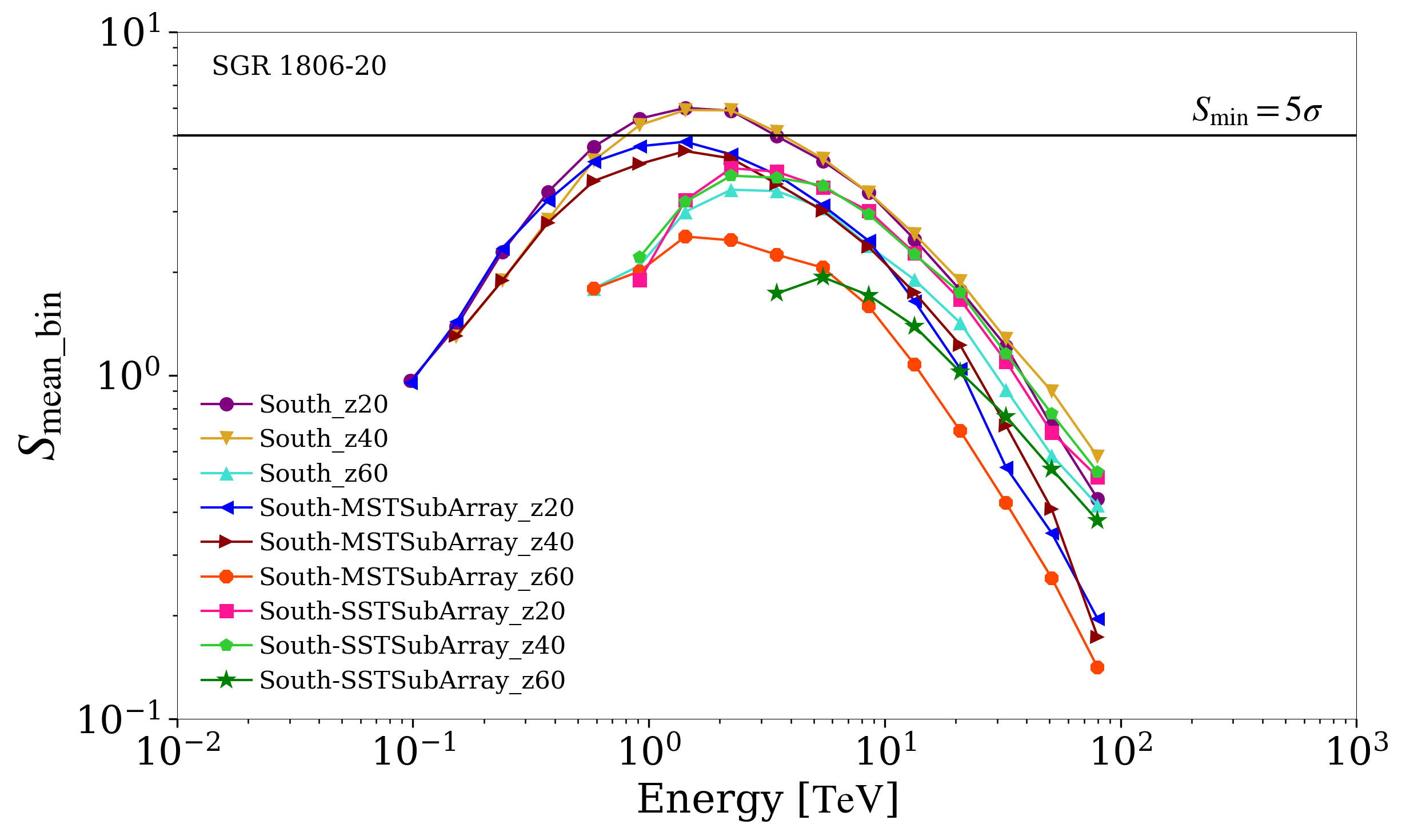}
    \caption{Mean statistical significance per energy bin was calculated from 3.000 simulated CTAO observations for a 10-hour observation period. These calculations are performed specifically for the SGR~1806-20, taking into account the CTAO southern hemisphere arrays with different zenith angles ($20^{\circ}$, $40^{\circ}$, $60^{\circ}$). The horizontal solid line indicates the minimum significance threshold used in generating the sensitivity curves.}
    \label{fig:Signif_bin}
\end{figure}

One might wonder why the CTAO configuration with the highest statistical significance for detection in the simulated observations does not align with the highest sensitivity curve. It is important to note that, in the simulation approach, significance is calculated based on the total number of counts within the specified energy range, while generating sensitivity curves involves applying certain criteria, including a minimum significance requirement for each energy bin. As a result, the excess per bin and, consequently, the significance per bin differ between the two methods. In the former case, the significance value for each bin may fall below the minimum threshold established in the latter. For example, when calculating the mean significance per bin ($S_{\rm mean\_bin}$) for simulated observations of the SGR~1806-20 region, we observe that this source consistently produces values $S_{\rm mean\_bin}$ below $5 \, \sigma$ for a $t_{\rm obs} = 10$~hours, except for all Southern arrays at $z = 20^{\circ}$ and $z = 40^{\circ}$. These two configurations demonstrate a narrow energy range around $2$~TeV with $S_{\rm mean\_bin} > 5 \, \sigma$ (see Fig.~\ref{fig:Signif_bin}), correlating with the results shown in the sensitivity curves of the top right panel of Fig.~\ref{fig:sens_10h}. Thus, the apparent discrepancy between the highest detection statistical significance and the best sensitivity curve arises because, while the simulated observations consider detection significance based on all simulated signal counts across all energy bins, the sensitivity curves assess signal counts based on minimum criteria for each energy bin.

\begin{figure*}[htb]
\centering
\gridline{\fig{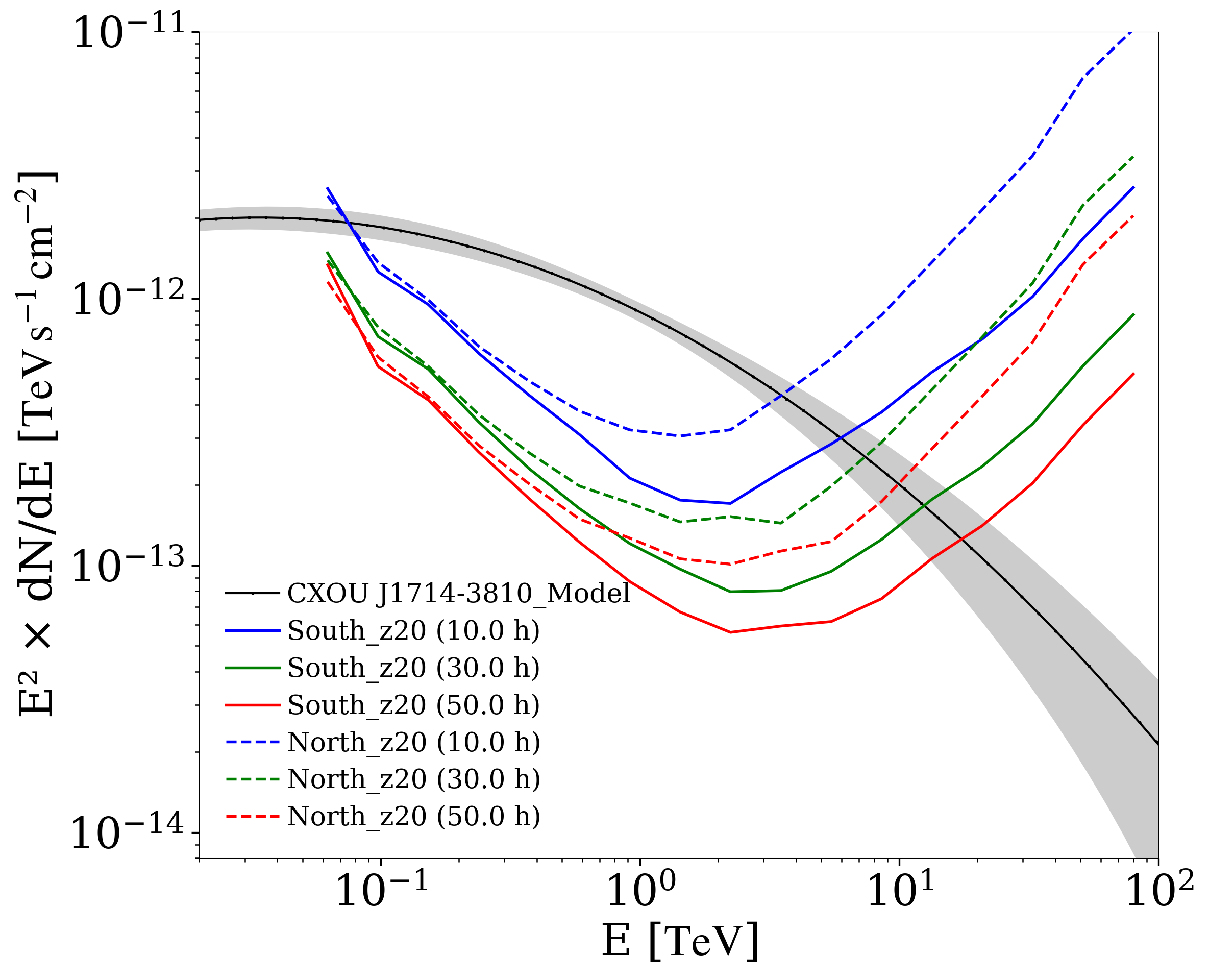}{0.33\textwidth}{(a) CXOU~J1714-3810 region}
          \fig{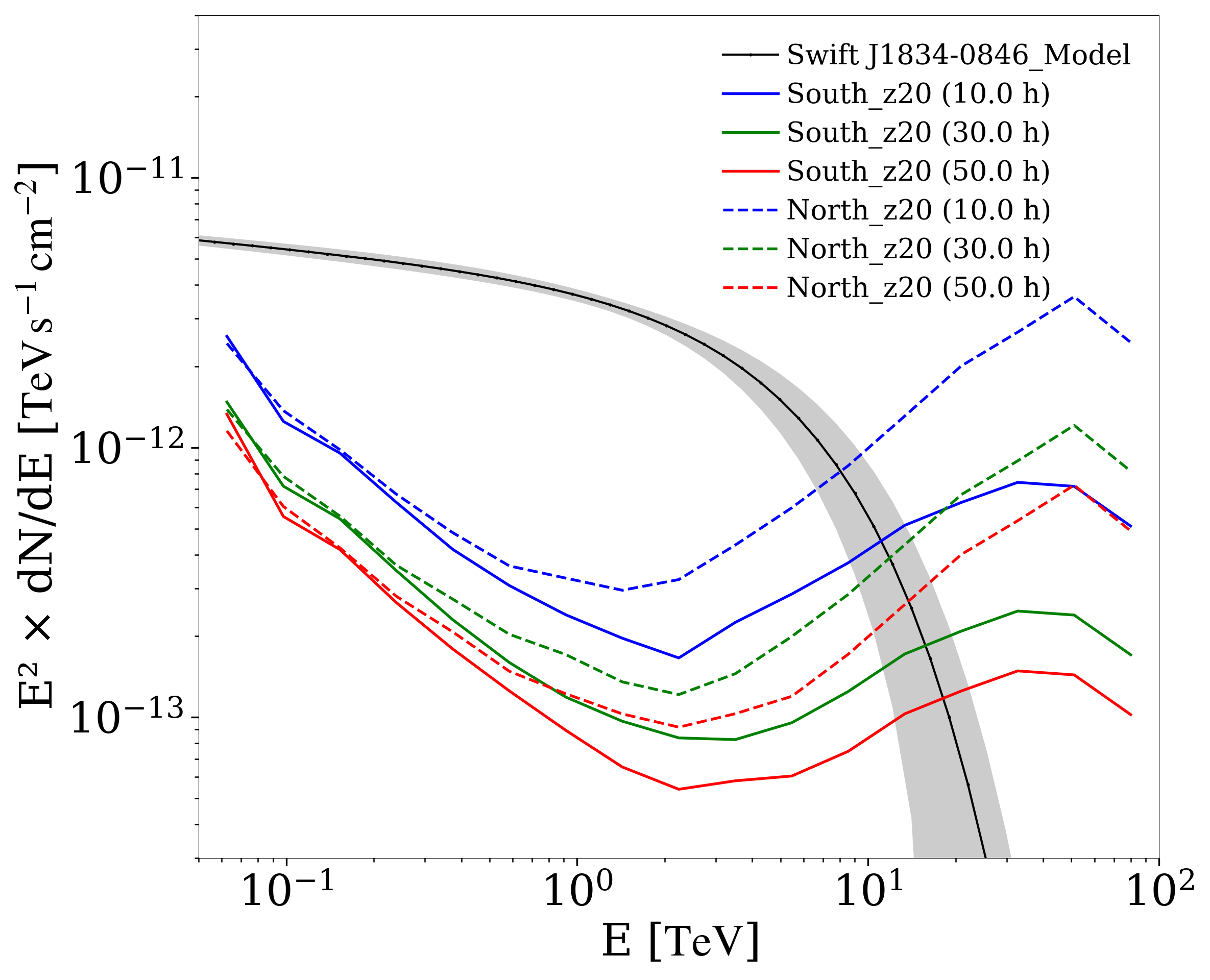}{0.33\textwidth}{(b) Swift~J1834-0846 region}
          \fig{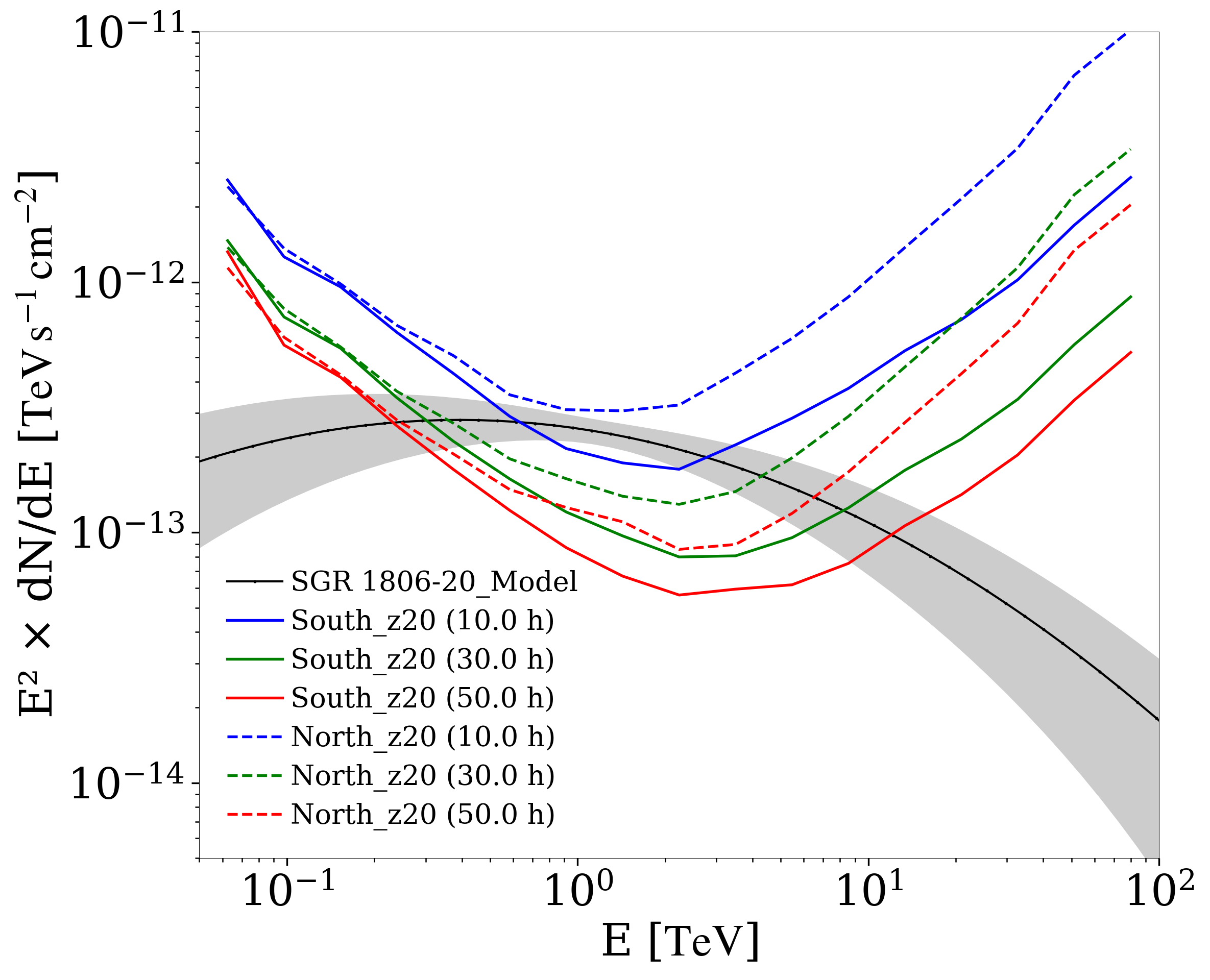}{0.33\textwidth}{(c) SGR~1806-20 region}}
\caption{The source spectral energy distribution and the differential sensitivity curves of the CTAO for the North and South full arrays with a zenith angle of $20^{\circ}$ are presented for three different observation times: $10$, $30$, and $50$ hours.}
\label{fig:sens_obstime}
\end{figure*}

Upon reviewing all sensitivity curves for both arrays and sub-arrays utilizing IRFs over a 10-hour period, it is evident that the IRFs of the CTAO South full array, especially at $z = 20^{\circ}$ and $z = 40^{\circ}$, exhibit superior performance in observing the three modeled sources. In terms of the CTAO's North location configuration, the IRFs of the full arrays at $z = 20^{\circ}$ and $z = 40^{\circ}$ demonstrate the highest sensitivity for detecting these sources. Moreover, the sensitivity curves presented in Fig.~\ref{fig:sens_10h} offer valuable insights into the most sensitive energy ranges for each set of IRFs, which can help prioritize specific configurations for observing targeted parts of the spectrum.

To evaluate the detectability characteristics of the modeled sources at different observation times, Fig.~\ref{fig:sens_obstime} displays the sensitivity curves for the full south and north arrays configured at $z = 20^{\circ}$. This analysis considers three observation times (10, 30, and 50 hours) along with the same detection criteria established earlier, including minimum significance, signal counts, and signal-to-background ratio. We selected these South and North IRFs because they represent the configurations with the highest sensitivity in detecting all three sources at each CTAO array location.

From Fig.~\ref{fig:sens_obstime}, it is clear that sensitivity improves with longer observation times. Comparing the sensitivity curves for 10h and 30h observations, we observe an enhancement in sensitivity ranging from approximately $1.7$ times at the lowest energies to 3 times at the highest energies of the CTAO. Similarly, the comparison between the 10- and 50-h curves shows an improvement in sensitivity by a factor of about 2 to 5 across the energy spectrum. Furthermore, it is evident that longer observation times are critical for detecting the most energetic spectral components of the CXOU~J1714-3810 and Swift~J1834-0846 regions. In the case of the SGR~1806-20 region, a 10-hour observation may be insufficient to detect this source given the previously established detection criteria. Extending the observation period to 30h or 50h, especially with the South full array, seems essential for capturing a wider range of this source's spectrum.

\begin{figure*}
\centering
    \gridline{\fig{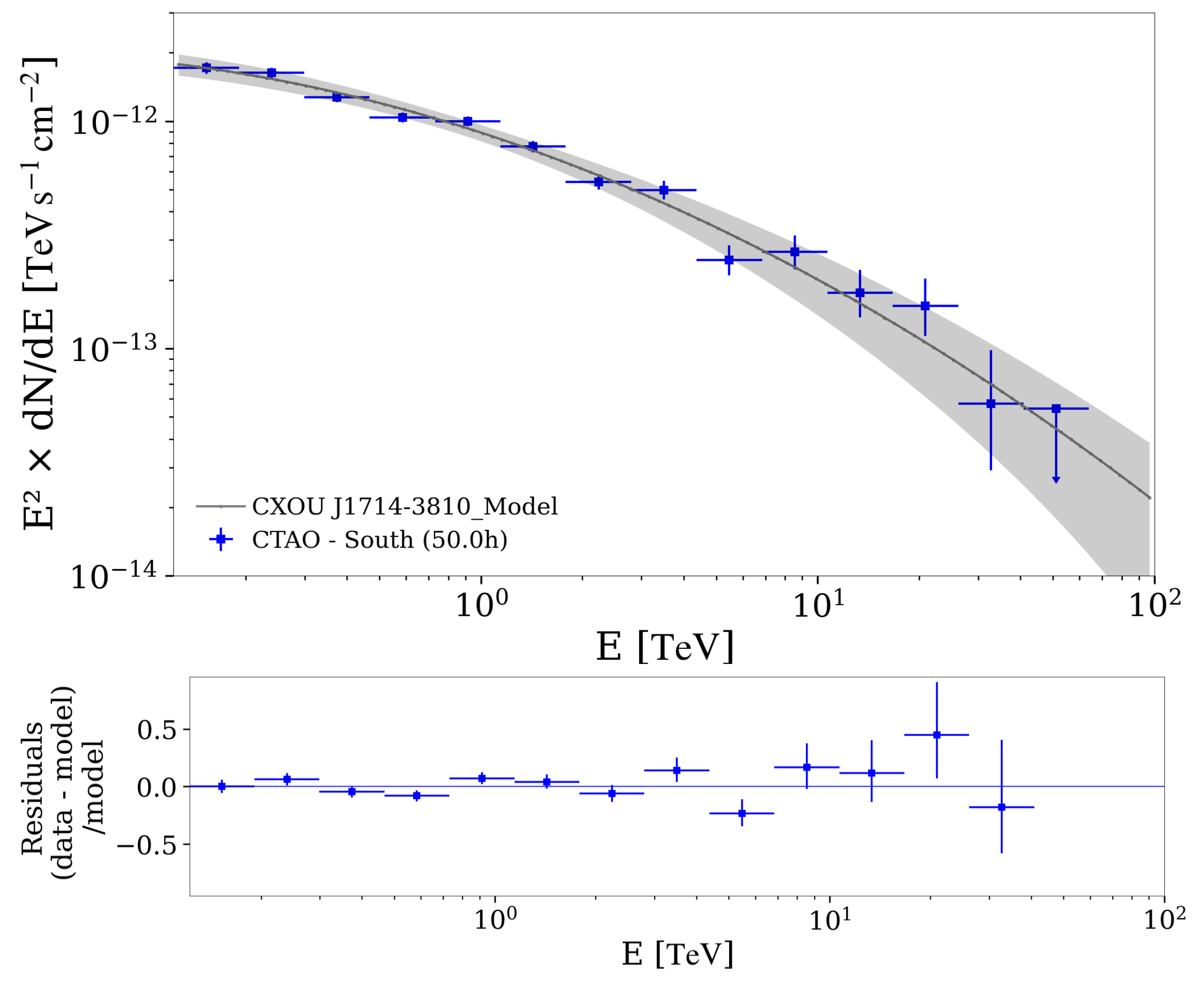}{0.33\textwidth}{(a) CXOU~J1714-3810 region}
          \fig{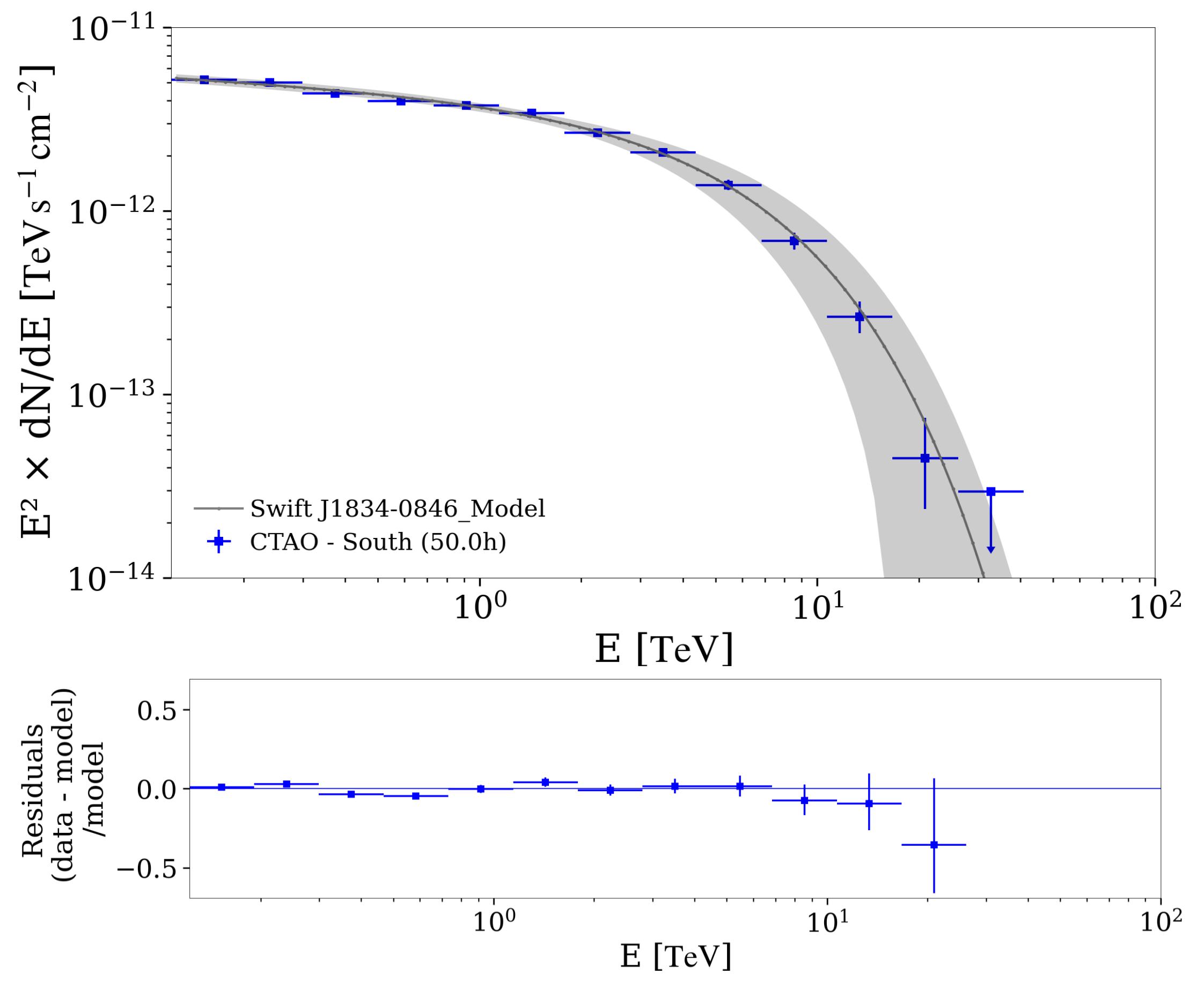}{0.33\textwidth}{(b) Swift~J1834-0846 region}
          \fig{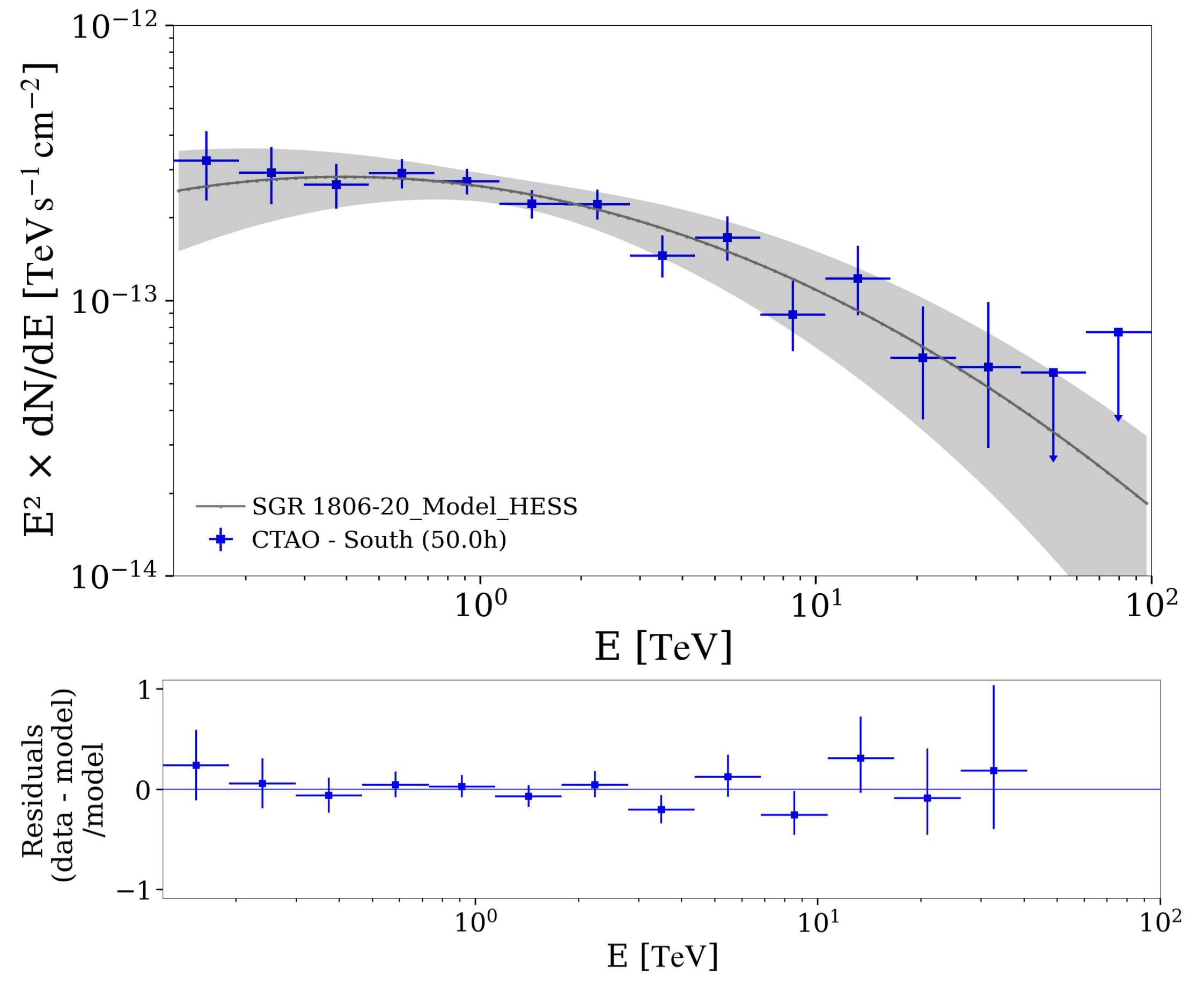}{0.33\textwidth}{(c) SGR~1806-20 region}}
    \caption{Simulated CTAO flux points and the source spectral energy distribution for the three magnetar regions. The flux points were estimated using the IRFs from the full Southern array, optimized for a zenith angle of $z = 20^{\circ}$ and an observation time of  $t_{\rm obs} = 50$~hours.}
    \label{fig:fluxp_50h}
    \vspace{0.3cm}
\end{figure*}

To further enhance our analysis, we estimated the flux points by fitting the source model across different energy bands. For this estimation, we selected one observation from the three thousand simulated CTAO observations for each modeled source, choosing one that exhibited a detection significance close to the mean significance of the entire set. Additionally, we considered a simulated observation using the IRFs from the full Southern array, optimized for a zenith angle of $z = 20^{\circ}$ with an observation time of $t_{\rm obs} = 50$~hours. The flux points were derived within the energy range of $100$~GeV to $100$~TeV, which is narrower than the expected energy range for CTAO in full-array configurations. Figure~\ref{fig:fluxp_50h} displays the results of this analysis for the three magnetar regions alongside the corresponding source model (see Table \ref{tab:Spec_model}).

When comparing the estimated flux points with the source models, we observe that the flux points closely follow the model curve, with residuals ([data-model]/model) up to $|0.5|$ for CXOU~J1714-3810 and Swift~J1834-0846, and up to $|1.0|$ for SGR~1806-20 (see Fig.~\ref{fig:fluxp_50h}). This indicates that with $50$~hours of observation, the CTAO will be able to reconstruct the actual behavior of the source emission curve, especially for CXOU~J1714-3810 and Swift~J1834-0846, which show a higher emission flux compared to SGR~1806-20.

\begin{figure*}[htb]
\centering
\gridline{\fig{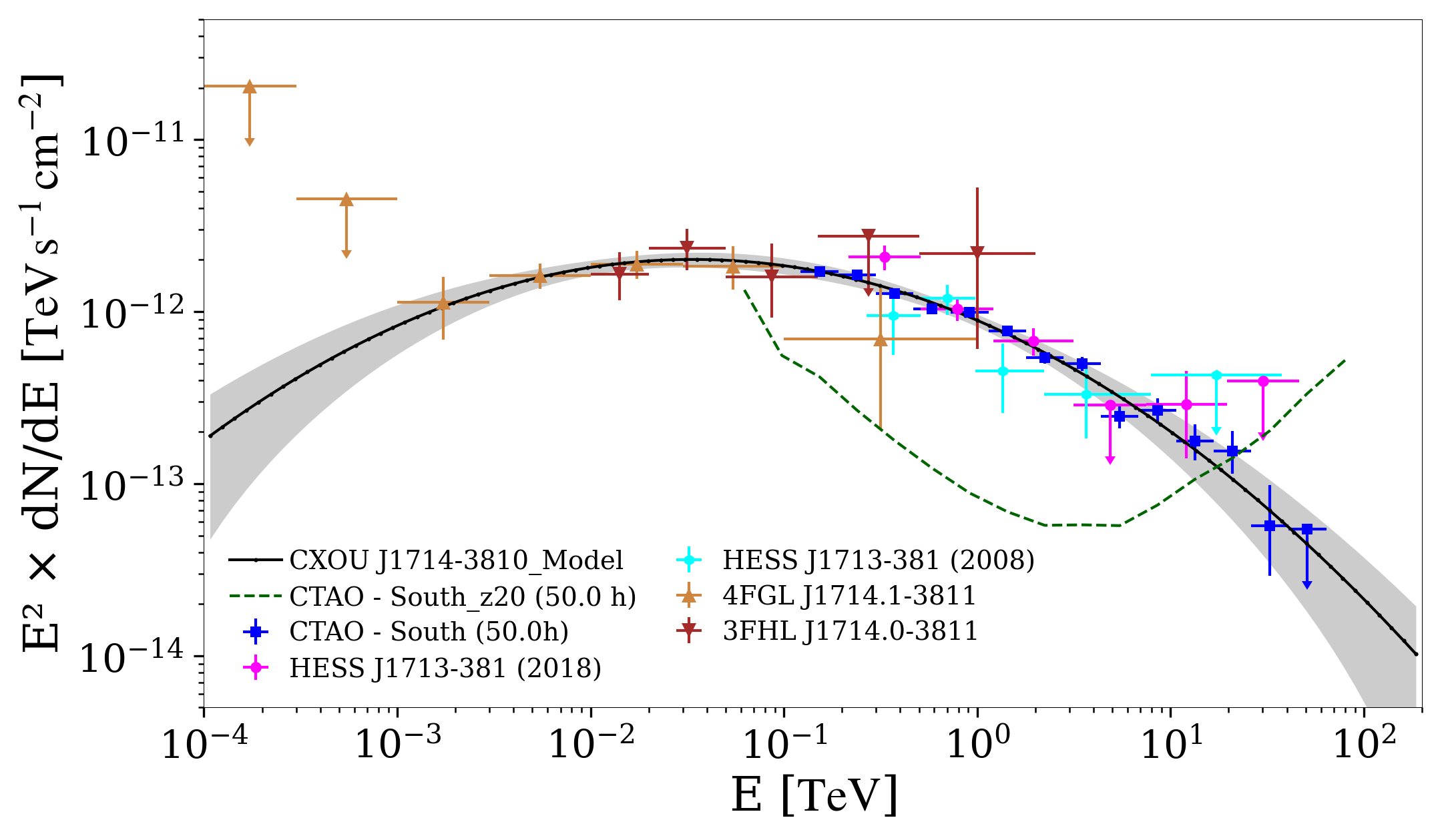}{0.5\textwidth}{(a) CXOU~J1714-3810 region}
          \fig{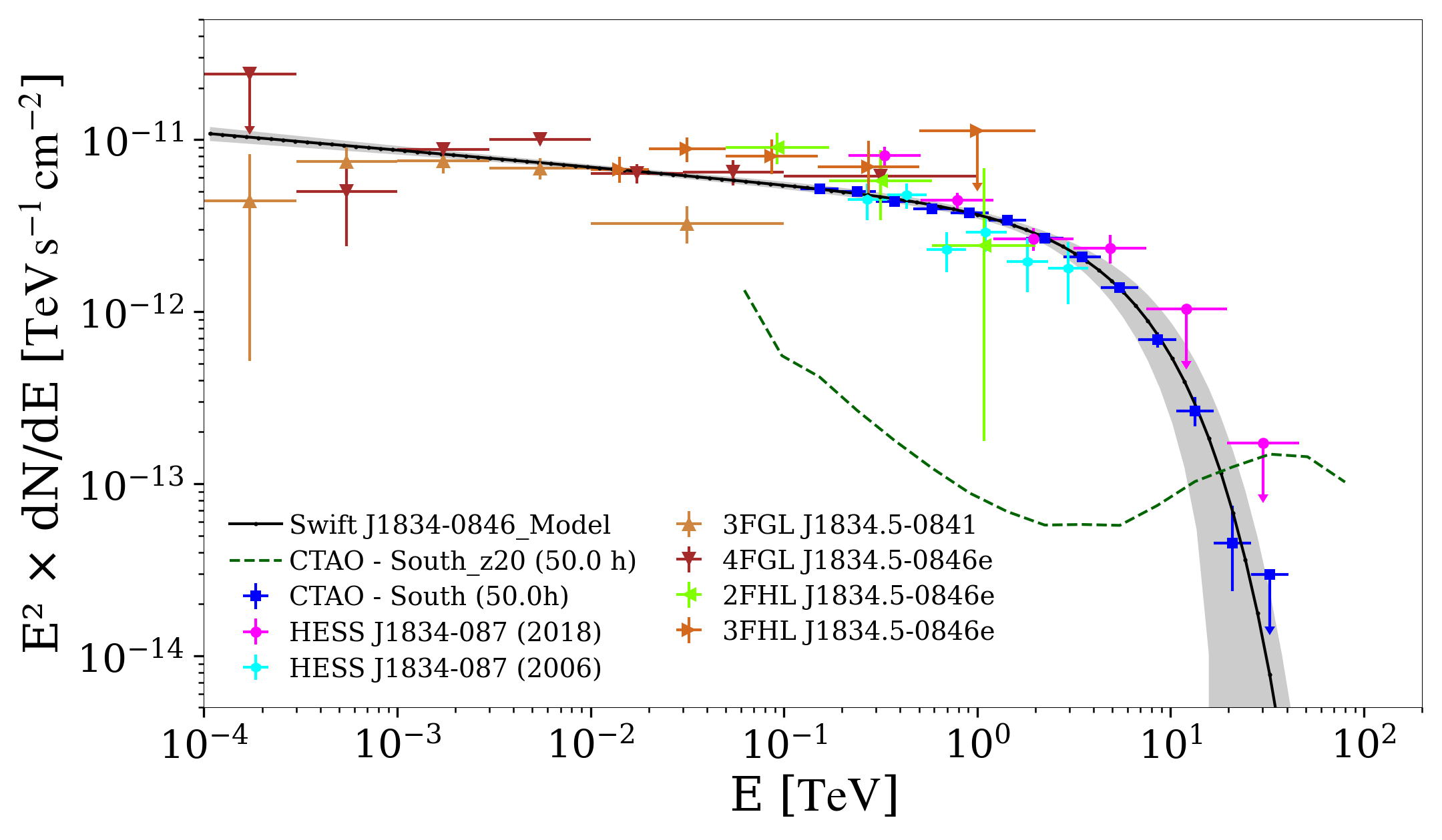}{0.5\textwidth}{(b) Swift~J1834-0846 region}}
\gridline{\fig{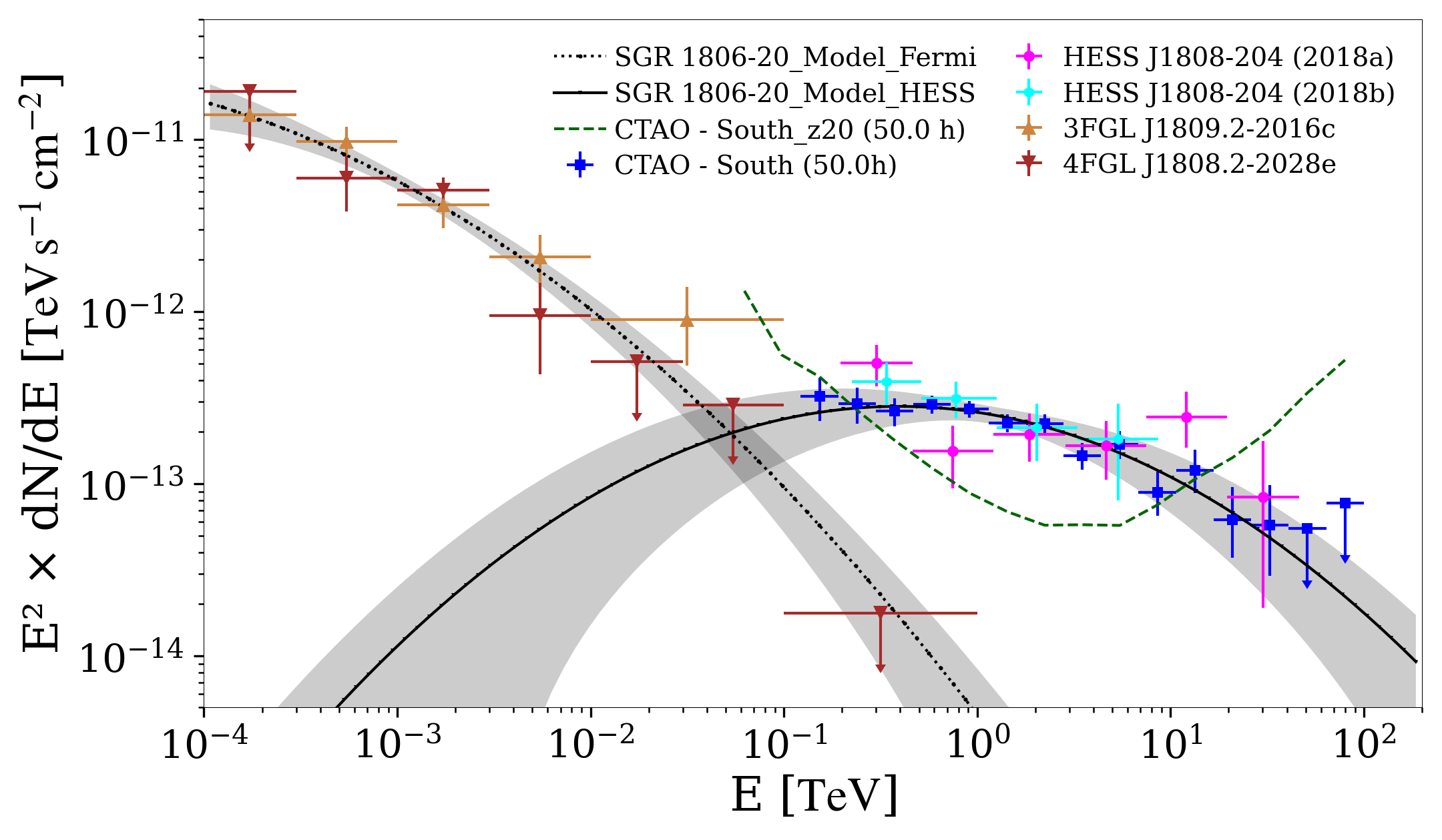}{0.5\textwidth}{(c) SGR~1806-20 region}}
\caption{The spectral energy distribution of the magnetar regions is presented alongside previous observations of counterparts included in the likelihood fit analysis, as described in Table~\ref{tab:counterparts}. The estimated flux points for the CTAO, derived from the observation simulation, are marked by blue squares. The solid line represents the spectral model obtained from the likelihood analysis, with the associated parameters specified in Table \ref{tab:Spec_model}. The dashed line illustrates the sensitivity curve for the IRFs of the Southern array at a zenith angle of $20^{\circ}$, recorded over a $50$-hour observation period. }
\label{fig:p_flux}
\end{figure*}

Figure \ref{fig:p_flux} presents the simulated CTAO flux points for the three magnetar regions, along with previous observations of potential counterparts identified in these areas (see Table \ref{tab:counterparts}). Additionally, Fig.~\ref{fig:p_flux} includes the sensitivity curve corresponding to the IRFs used in the flux point estimation, along with the spectral models from Table \ref{tab:Spec_model} obtained through simultaneous likelihood fitting.

Regarding the flux points for CXOU~J1714-3810 and Swift~J1834-0846 regions, it is evident that the energy flux errors for CTAO are very small, particularly at lower energies, confirming that CTAO will detect these regions with significantly greater precision than previous observatories. In fact, when comparing the flux points estimated by CTAO with those measured by H.E.S.S. (Livetime $\sim 100$~h, $44$~h for HESS~J1713-381 and HESS~J1834-087, respectively; see \citet{2018A&A...612A...1H}), we find that the error bars improve by at least a factor of $3$ for lower energies and by at least a factor of $4$ for energies above $10$~TeV.

For the SGR~1806-20 region, the flux errors are larger compared to CXOU~J1714-3810 and Swift~J1834-0846 regions due to its lower flux. However, even here, the CTAO reduces the errors significantly compared to the flux points measured by H.E.S.S. (Livetime $\sim 80$~h for HESS~J1808-204; see \citet{2018A&A...612A...1H}). In this case, error bars decrease by at least a factor of $2$ for energies below $10$~TeV and by a factor of $2.5$ for higher energies.

The increase in energy flux error with energy is related to the number of counts detected above the background in each energy bin, demonstrating that the photon excess for these regions decreases substantially in CTAO's higher energy range.

Furthermore, by analyzing the CTAO flux points and the sensitivity curve, we find that the CTAO observations of the CXOU~J1714-3810 and Swift~J1834-0846 magnetar regions offer improved spectral resolution in the energy range around $10$~TeV compared to previous observations by H.E.S.S., which reported flux points as upper limit values. CTAO's detection of these regions is expected to provide a more precise observation of energy flux decay, thereby refining constraints on the spectral model parameters. Additionally, observation of the SGR~1806-20 region will enhance the differentiation of the spectral model parameters that describe this source and better constrain the model's curvature, which is inadequate when evaluated solely with the flux points from H.E.S.S. The 50-hour sensitivity curve for this region suggests that flux points above $10$~TeV will fall below the established minimum detection criteria. Therefore, to achieve a more comprehensive description of this source, it may be necessary to extend the observation time or adjust the minimum detection criteria.

Our analysis of the CTAO flux points from a 50-hour observation confirms the observatory's capability to detect counts in the energy range up to approximately $50$~TeV for the three modeled sources. However, it's important to highlight that not all of these flux points are detectable according to the 50-hour sensitivity curves. This discrepancy arises because the flux points are generated from simulations of CTAO observations, leading to detection criteria that differ from those used to create the sensitivity curves. As a result, CTAO flux points that fall below the sensitivity curve have a significance $S_{\rm bin} < 5 \sigma$, while upper limits are established in energy bins where the significance is $S_{\rm bin}  \leqslant 2 \sigma$.

Essentially, the sensitivity curves reflect the observatory's detectability within the constraints established by significance, expected signal counts, and excess-criteria that are not fulfilled by the flux points positioned below the curve. Nevertheless, even though these flux points do not statistically satisfy the detection criteria, they still represent a detectable number of photon counts above the background from the source regions, albeit with lower significance. These flux points are valuable as they provide insight into the overall behavior of the emission curve (See Fig.~\ref{fig:p_flux}).

\section{Summary and conclusions}\label{sec:6}

In this article, we have investigated the regions surrounding three magnetars—CXOU~J1714-3810, Swift~J1834-0846, and SGR~1806-20—that are known to emit electromagnetic signals previously detected by the Fermi-LAT and H.E.S.S. telescopes. We assess the capability of the CTAO to detect gamma-ray emissions from these regions by conducting an ON/OFF spectral analysis with the Gammapy software. Our analysis uses a simulation-based approach to predict gamma-ray emissions from selected magnetar regions. We employed a joint likelihood fit using data from the Fermi-LAT and H.E.S.S. instruments to establish the source model across the multi-GeV to multi-TeV energy range. Based on this model, we simulated the expected gamma-ray flux using the 1D ON/OFF observation approach. Furthermore, we compared the sensitivity of CTAO against the simulated flux points and existing observational data, emphasizing the advancements in detection capabilities.

We investigated the simulated CTAO detections and their average significance $S_{\rm mean}$, finding that both the full southern and northern CTAO arrays can observe the CXOU~J1714-3810 and Swift~J1834-0846 magnetar regions with a $S_{\rm mean}$ exceeding $10$ and $30 \, \sigma$, respectively, within just $5$~hours of observation. In contrast, the SGR~1806-20 region requires an observation time of at least $20-30$~hours to achieve a $S_{\rm mean} \gtrsim 5 \, \sigma$.

The simulation results also show that CTAO will detect gamma-ray emissions from the three magnetar regions with much lower errors in emission flux compared to current instruments, especially in the energy range above $1-10$~TeV. The study of CTAO flux points and sensitivity curves demonstrates that CTAO's observations of the magnetar regions CXOU J1714-3810 and Swift J1834-0846 will offer enhanced spectral resolution around $10$~TeV, surpassing previous H.E.S.S. observations that only provided upper limit flux values. As a result, CTAO is expected to deliver more accurate measurements of energy flux decay, leading to refined constraints on the spectral model parameters. Additionally, CTAO's observations of the SGR 1806-20 region will help better define the model's curvature, which could not be fully assessed using H.E.S.S. flux points alone. These results suggest that CTAO will be pivotal in identifying and characterizing the particle acceleration processes occurring in these magnetar environments.

Furthermore, our analysis of CTAO flux points from a 50-hour observation confirms the observatory's ability to detect counts up to around $50$~TeV for the three modeled sources. However, not all flux points align with the 50-hour sensitivity curves due to differences between the simulation-based detection criteria and those used to generate the curves. The sensitivity curves reflect detectability based on significance, expected signal counts, and excess criteria, which are not met by flux points below the curve. Despite not meeting strict statistical detection thresholds, these lower-significance flux points are still detectable by CTAO and may provide valuable insight into the emission curve's behavior.

Our analysis of the CTAO’s potential to detect gamma-ray emissions from CXOU~J1714-3810, Swift~J1834-0846, and SGR~1806-20 underscores the observatory's critical role in advancing our understanding of cosmic-ray acceleration in the regions surrounding magnetars. The results suggest that extended observation times or adjustments to detection criteria could yield even more detailed insights. Future research should focus on refining the observational strategies to enhance detection sensitivity, potentially unlocking new information about the particle acceleration mechanisms in these powerful astrophysical sources.




\vspace{0.8cm}
{\it{Acknowledgments:} We thank the anonymous referee for their valuable suggestions and comments, which have helped improve this work. M.F.S., R.P.C., J.G.C. and R.C.A. acknowledge the financial support of the NAPI “Fenômenos Extremos do Universo” of Fundação de Apoio à Ciência, Tecnologia e Inovação do Paraná. M.F.S. thanks the support of Conselho Nacional de Desenvolvimento Cient\'{i}fico e Tecnol\'{o}gico (CNPq) (173535/2023-2). J.G.C. is grateful for the support of Funda\c c\~ao de Amparo \`{a} Pesquisa e Inova\c c\~ao do Esp\'irito Santo (FAPES) (grant Nos. 1020/2022, 1081/2022, 976/2022, 332/2023), CNPq (grant No. 311758/2021-5), and Funda\c c\~ao de Amparo \`{a} Pesquisa do Estado de S\~ao Paulo (FAPESP) (grant N. 2021/01089-1). R.C.A. research is supported by CAPES/Alexander von Humboldt Program (88881.800216/2022-01); CNPq (310448/2021-2) and (4000045/2023-0); Araucária Foundation (698/2022) and (721/2022); and FAPESP (2021/01089-1). R.C.A. also acknowledges the support of L’Oreal Brazil, with the partnership of ABC and UNESCO in Brazil. The authors acknowledge the AWS Cloud Credit/CNPq and the National Laboratory for Scientific Computing (LNCC/MCTI, Brazil) for providing HPC resources of the SDumont supercomputer, which have contributed to the research results reported in this paper. URL: https://sdumont.lncc.br.}



\bibliography{sample631,references}
\bibliographystyle{aasjournal}



\end{document}